\documentclass[journal]{IEEEtran}
\usepackage{amsfonts}
\usepackage{amssymb}
\usepackage{amsthm}
\usepackage{amsmath,amsfonts,amssymb}
\usepackage[dvips]{graphicx}
\usepackage{verbatim}
\usepackage{setspace}
\usepackage{bm}
\usepackage{subfigure}
\usepackage{algorithmic} 
\usepackage[ruled,vlined]{algorithm2e}
\usepackage{cite}
\usepackage{enumerate}
\usepackage{indentfirst}
\usepackage{changepage}
\usepackage{pdfpages}
\usepackage{color}
\usepackage{lettrine}

\newtheorem{criterion}{Criterion}
\begin{document}
	
\title{Channel Estimation for Movable Antenna Communication Systems: A Framework Based on Compressed Sensing}

\author{
	Zhenyu Xiao,~\IEEEmembership{Senior Member,~IEEE,}
	Songqi Cao,
	Lipeng Zhu,~\IEEEmembership{Member,~IEEE,}
	Yanming Liu,~\IEEEmembership{Graduate Student Member,~IEEE,}
	Xiang-Gen Xia,~\IEEEmembership{Fellow,~IEEE,}
	and Rui Zhang~\IEEEmembership{Fellow,~IEEE}
	
	\thanks{This work was supported in part by the National Natural Science Foundation
		of China (NSFC) under grant numbers 62171010 and U22A2007.}
	\thanks{Z.~Xiao, S.~Cao and Y.~Liu are with the School of Electronic and Information Engineering, Beihang University, Beijing 100191, China. (e-mail: xiaozy@buaa.edu.cn, csq@buaa.edu.cn, liuyanming@buaa.edu.cn).}
	\thanks{L.~Zhu is with the Department of Electrical and Computer Engineering, National University of Singapore, Singapore 117583, Singapore. (e-mail: zhulp@nus.edu.sg).}
	\thanks{X.-G. Xia is with the Department of Electrical and Computer Engineering, University of Delaware, Newark, DE 19716, USA. (e-mail: xxia@ee.udel.edu).}
	\thanks{R. Zhang is with School of Science and Engineering, Shenzhen Research Institute of Big Data, The Chinese University of
		Hong Kong, Shenzhen, Guangdong 518172, China (e-mail:rzhang@cuhk.edu.cn). He is also with the Department of Electrical
		and Computer Engineering, National University of Singapore, Singapore 117583 (e-mail: elezhang@nus.edu.sg).}
}	

\maketitle

\begin{abstract}
	Movable antenna (MA) is a new technology with great potential to improve communication performance by enabling local movement of antennas for pursuing better channel conditions. In particular, the acquisition of complete channel state information (CSI) between the transmitter (Tx) and receiver (Rx) regions is an essential problem for MA systems to reap performance gains. In this paper, we propose a general channel estimation framework for MA systems by exploiting the multi-path field response channel structure. Specifically, the angles of departure (AoDs), angles of arrival (AoAs), and complex coefficients of the multi-path components (MPCs) are jointly estimated by employing the compressed sensing method, based on multiple channel measurements at designated positions of the Tx-MA and Rx-MA. Under this framework, the Tx-MA and Rx-MA measurement positions fundamentally determine the measurement matrix for compressed sensing, of which the mutual coherence is analyzed from the perspective of Fourier transform. Moreover, two criteria for MA measurement positions are provided to guarantee the successful recovery of MPCs. Then, we propose several MA measurement position setups and compare their performance. Finally, comprehensive simulation results show that the proposed framework is able to estimate the complete CSI between the Tx and Rx regions with a high accuracy. 
\end{abstract}


\section{Introduction}\label{section_1}

With the explosive growth of wireless applications, multiple-input multiple-output (MIMO) technologies have been widely investigated and utilized in existing wireless communication systems because of their ability of exploiting new degrees of freedom (DoFs) in the spatial domain \cite{telatar1999capacity,paulraj2004overview,stuber2004broadband,6798744,8789678}. MIMO technologies not only bring beamforming gains and spatial multiplexing gains to increase the transmission rate, but also  provide spatial diversity to enhance the reliability of wireless communications \cite{9348800,8804165,1519678}. However, conventional MIMO antennas are discretely placed at fixed positions in general, and thus the spatial DoFs cannot be fully utilized as the wireless channel variation in continuous spatial regions is not fully exploited by conventional MIMO systems. 

To fully explore the spatial DoFs, a novel antenna architecture, namely movable antenna (MA), was proposed \cite{zhu2022modeling}. MAs can fully exploit spatial diversity by flexibly adjusting their positions in the local transmitter (Tx)/receiver (Rx) region. Specifically, due to the superposition of multi-path components, the channel gain between the Tx and Rx varies with the antennas' locations in the spatial region, which is known as small-scale fading. By continuously moving the MAs in the spatial region, the MA system can achieve the maximum channel gain between the Tx and Rx, thereby obtaining a much higher receive signal-to-noise ratio (SNR) compared to conventional fixed-position antennas (FPAs) which may experience deep fading channels. Moreover, by moving the MAs to the positions with minimum channel gain with respect to the jammer, the interference can be efficiently mitigated. In addition to improving the channel power gain, the MA system provides additional DoFs in beamforming and spatial multiplexing\cite{zhu2023mov}. In particular, more flexible beamforming can be achieved by jointly designing the positions of the MAs and beamforming weights. For spatial multiplexing, the channel matrix of the MA-MIMO system can be reshaped by optimizing the positions of MAs, and thus the channel capacity can be further increased compared to conventional FPA-MIMO systems \cite{ma2022mimo}.

There have been some preliminary studies demonstrating the potentials of MAs in wireless communications \cite{ma2022mimo,zhu2023movable,xiao2023multiuser,zhu2023array,zhu2022modeling,zhu2023mov,wu2023movable,cheng2023sum,chen2023joint,cheng2023movable}. The concept of MA was first proposed in \cite{zhu2022modeling}, where a field-response channel model was developed to characterize the variation of the channel gain with respect to the positions of the MAs. Then, based on this new channel model, the authors analyzed the maximum channel gain obtained by a single MA and compared it to conventional FPA systems. It was demonstrated that MA systems can achieve significant performance gains over conventional FPA systems, as well as achieve comparable performance to single-input multiple-output (SIMO) beamforming systems. In addition, the potentials and advantages of MA were analyzed in \cite{zhu2023mov}, from the perspective of signal power improvement, interference mitigation, flexible beamforming, and spatial multiplexing. Moreover, an overview of the application scenarios for MA systems was illuminated, which indicated that MAs are promising for machine-type communications with slowly-varying channels as well as satellite communication for their capability in flexible beamforming. Furthermore, the channel capacity for MA-aided MIMO systems was maximized in \cite{ma2022mimo} by jointly optimizing the positions of MAs and the covariance matrix of transmit signals. Compared to conventional FPA systems, the channel matrix for MA systems can be reshaped by exploiting MA position optimization, and thus the channel capacity can be improved. In \cite{zhu2023movable}, an MA-aided multiuser communication system was considered. The total transmit power of MA-aided users was minimized by jointly optimizing the positions of the MAs at the user side, the transmit power of users, and the receive combining matrix at the BS under the constraint of minimum achievable rate for each user. In comparison, the position optimization for multiple MAs at the base station (BS) was investigated in \cite{xiao2023multiuser}, where the minimum achievable rate of multiple users was maximized by jointly designing the MA positioning and receive combining at the BS. The MA-enhanced beamforming was investigated in \cite{zhu2023array}, which showed interestingly that the full array gain over the desired signal direction and the null steering over undesired interference directions can be concurrently achieved by MA arrays. The authors in \cite{wu2023movable} considered the joint optimization of beamforming and MAs' discrete positions at the BS for downlink multiuser multiple-input single-output (MISO) systems, where optimal solutions for minimizing the total transmit power were obtained under the constraint of the minimum required signal-to-interference-plus-noise ratio (SINR) of each individual user. Moreover, the joint optimization of the Tx beamforming vector and Tx-MAs' positions was investigated in \cite{cheng2023sum} to maximum the sum-rate for downlink multiuser MISO transmission. In \cite{chen2023joint}, a point-to-point MA-enabled MIMO system was considered to maximize the ergodic achievable rate by jointly optimizing the MA positions and the transmit covariance matrix based on statistical channel state information (CSI). The authors in \cite{cheng2023movable} extended the application of MAs to over-the-air computation (AirComp) systems, where the computational mean-squared error (MSE) was minimized by jointly optimizing the transmit power, MAs' positions, and receive combining.


Although MA systems have many advantages over conventional FPA systems, the performance improvement relies on knowing the accurate CSI between the Tx and Rx. For conventional FPA communication systems, e.g., in \cite{7458188, li2017millimeter,gao2015spatially}, the channel estimation is only performed at finite positions where the antennas are located. In comparison, for finding the best MAs' positions, MA-aided communication systems need to reconstruct the complete channel response between any point in the Tx region and any point in the Rx region, which contains an infinite number of positions due to the continuity of the spatial regions. Traversing all the positions in the Tx and Rx regions for channel measurement requires extremely high pilot overhead and time consumption, which is infeasible in practice. Thus, it is expected to find a new strategy for reconstructing the complete CSI between the Tx and Rx regions by conducting channel measurement at a finite number of MA positions. 

Given the above considerations, a successive transmitter-receiver compressed sensing (STRCS) method was proposed for MA channel estimation in \cite{10236898}. Specifically, the angles of departure (AoDs), angles of arrival (AoAs), and complex coefficients of the multi-path components (MPCs) were sequentially estimated, based on which the channel response between the entire Tx region and the entire Rx region was reconstructed. However, the sequential estimation of the MPCs' information may lead to a cumulative error and require high channel measurement overhead. To overcome this limitation, in this paper, we study the channel estimation framework based on compressed sensing for MA-aided communication systems. By jointly estimating the MPCs' information, our proposed method can significantly estimate/reconstruct the channel response between the entire Tx region and the entire Rx region via a smaller number of channel measurements, and the estimation error can be significantly reduced compared to \cite{10236898}. The main contributions of this paper are summarized as follows:
\begin{itemize}
	\item A generic channel estimation framework for MA systems is proposed. Specifically, based on the field-response channel model \cite{zhu2022modeling}, the channel reconstruction can be implemented by estimating the AoDs, AoAs, and complex coefficients of the MPCs between the Tx and Rx regions. By discretizing the AoDs and AoAs, we formulate the channel estimation as a sparse signal recovery problem, where the MPCs are estimated by the compressed sensing method via the orthogonal matching pursuit (OMP) algorithm.
	\item In the proposed channel estimation framework, the measurement matrix is fundamentally determined by the MA measurement positions, which significantly impact the channel estimation performance. Thus, we analyze the mutual coherence of the measurement matrix from the perspective of Fourier transform. Two criteria for MA measurement positions are derived to guarantee successfully resolving the MPCs in the angular domain. Accordingly, five setups for MA measurement with deterministic or random antenna positions are proposed. Moreover, we evaluate the mutual coherence of the measurement matrices for the proposed MA measurement setups and compare their performance in channel estimation via simulations.
	\item Numerical results are provided to evaluate the accuracy of channel estimation for the proposed compressed sensing-based framework. It is shown that the proposed solutions can obtain the complete CSI between the entire Tx and Rx regions with a high accuracy. Moreover, the receive SNR achieved by the considered MA system via antenna position optimization under the estimated CSI is almost the same as that under perfect CSI. 
\end{itemize}

The rest of this paper is organized as follows. Section \ref{section_2} introduces the signal model and field-response channel model of the MA system. In Section \ref{section_3}, the proposed compressed sensing-based channel estimation framework for MA communication systems is presented. In Section \ref{section_4}, we analyze the mutual coherence of the measurement matrix from the perspective of Fourier transform and propose several MA measurement position setups. In Section \ref{section_5}, numerical results are provided to verify the efficacy of the proposed strategy. Finally, we conclude this paper in Section \ref{section_6}. 

\begin{figure*}[h]
	\centering
	\includegraphics[width=14 cm]{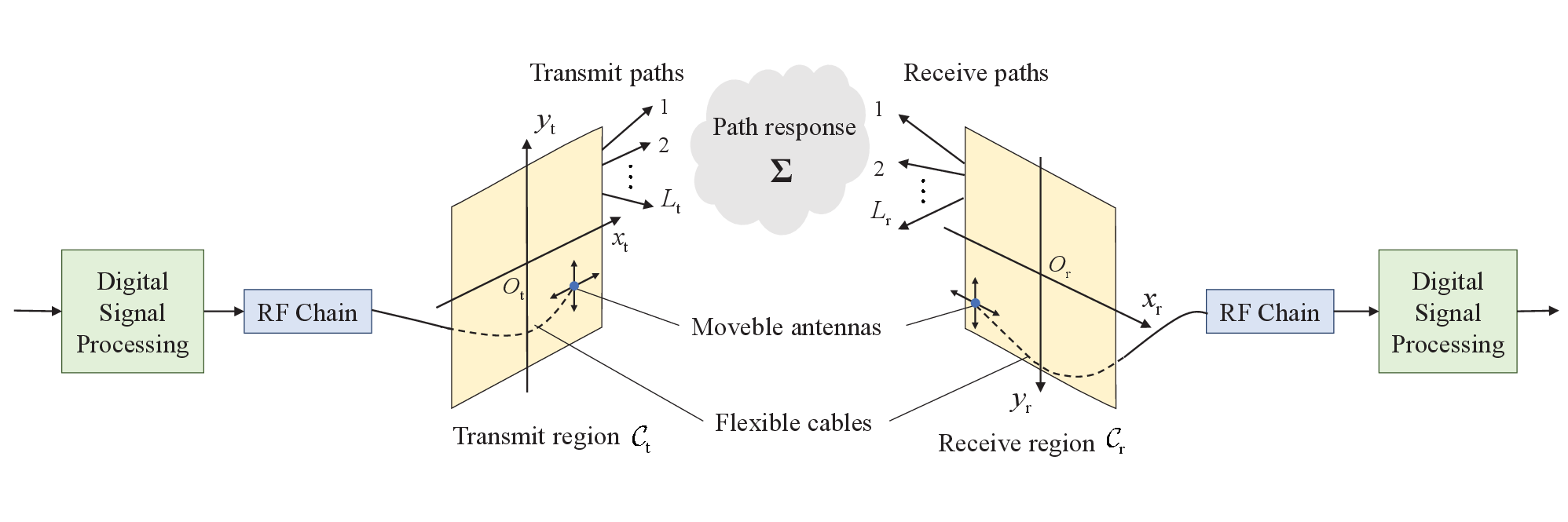}
	\caption{Illustration of the MA-aided communication system, where the Tx-MA and Rx-MA are connected to the radio frequency (RF) chains via flexible cables to enable local movement.}
	\label{fig:system}
\end{figure*}

\emph{Notation}: $a, {\bf a}, {\bf A}$, and $\mathcal{A}$ denote a scalar, a vector, a matrix, and a set, respectively. $\left[{\bf A}\right]_{i,:}$, $\left[{\bf A}\right]_{:,j}$, and $\left[{\bf A}\right]_{i,j}$ represent the $i$-th row, the $j$-th column, and the $i$-th row and $j$-th column entry of matrix ${\bf A}$, respectively. $\left(\cdot\right)^{\rm T}$ and $\left(\cdot\right)^{\rm H}$ denote transpose and conjugate transpose, respectively. $\lfloor\cdot\rfloor$ and $\lceil \cdot \rceil$ represent the floor and ceiling operations, respectively. Denote $\bmod$ and ${\rm vec}$ as the modulo operation and vectorization, respectively. In addition, $\otimes$ denotes the Kronecker product. $\mathcal{CN}\left(0,\delta^2\right)$ represents the circularly symmetric complex Gaussian (CSCG) distribution with mean zero and variance $\delta^2$. $\mathbb{C}$ and $\mathbb{Z}$ denote the sets of complex numbers and integers, respectively. $\vert \cdot \vert, \Vert \cdot \Vert_0$, and $\Vert \cdot \Vert_2$ denote the absolute value, $l_0$-norm, and $l_2$-norm, respectively. ${\bf 1}_{N\times M}$ denotes the $N\times M$ matrix with all entries equal to $1$. 

\section{System and Channel Model}\label{section_2}

The architecture of the considered MA-aided communication system is shown in Fig. \ref{fig:system}, where a single Tx-MA and a single Rx-MA are employed at the Tx and Rx, respectively\footnote{Note that the proposed channel estimation framework is also applicable to multi-MA systems, where the multiple Tx-/Rx-MAs can be simultaneously moved for channel measurements at different positions.}. Local Cartesian coordinate systems, $x_{\rm t}\text{-}O_{\rm t}\text{-}y_{\rm t}$ and $x_{\rm r}\text{-}O_{\rm r}\text{-}y_{\rm r}$, are established to describe the positions of the MAs in the Tx and Rx regions, $\mathcal{C}_{\rm t}$ and $\mathcal{C}_{\rm r}$, respectively. For convenience, we assume that both the Tx and Rx regions are square regions of size $R\lambda\times R\lambda$. Denote ${\bf t}=\left[x_{\rm t}, y_{\rm t}\right]^{\rm T}$ and ${\bf r}=\left[x_{\rm r}, y_{\rm r}\right]^{\rm T}$ as the coordinates of the Tx-MA and Rx-MA, respectively. 

\begin{figure}[tbp]
	\centering
	\includegraphics[width=5.3 cm]{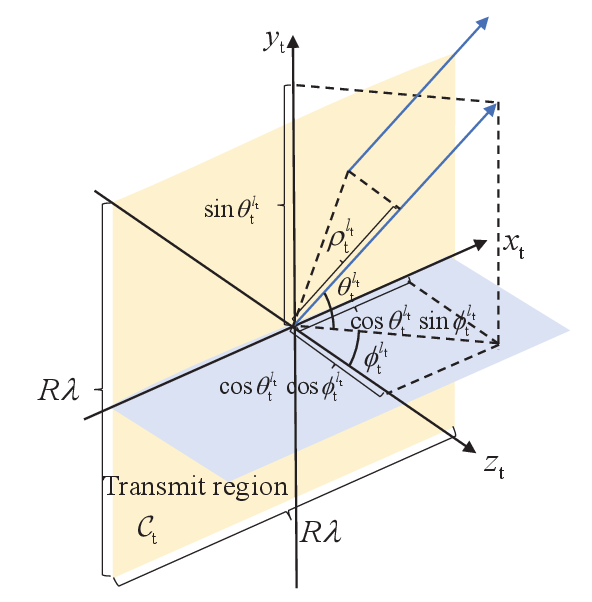}
	\caption{Illustration of the spatial angles and the signal propagation distance difference for the $l_{\rm t}$-th path in the Tx region.}
	\label{fig:ma_system}
\end{figure}

Denoting the transmit pilot signal as $s$, the received signal at the Rx can be expressed as
\begin{equation}\label{signal_model}
	v\left({\bf t, r}\right)=h\left({\bf t, r}\right)\sqrt{p_{\rm t}}s+z,
\end{equation}
where $h\left({\bf t, r}\right)$ represents the channel coefficient between ${\bf t}$ and ${\bf r}$,  $p_{\rm t}$ represents the transmit power at the Tx, and $z\in \mathcal{CN}\left(0, \delta^2\right)$ is the additive white Gaussian noise (AWGN) at the Rx with power $\delta^2$. Thus, the SNR for the received signal is given by
\begin{equation}\label{snr}
	\gamma\left({\bf t, r}\right)=\frac{\left|h\left({\bf t, r}\right)\right|^2p_{\rm t}}{\delta^2}.
\end{equation}

Next, we model the effect of the positions of the Tx-MA and Rx-MA on channel response $h\left({\bf t, r}\right)$. In general, the channel between the transceivers is the superposition of MPCs \cite{zhu2023movable}. Denote the number of transmit paths and receive paths as $L_{\rm t}$ and $L_{\rm r}$, respectively. Path-response matrix (PRM) ${\bf \Sigma}\in \mathbb{C}^{L_{\rm r}\times L_{\rm t}}$ is defined to represent the responses between all channel paths from transmit reference position ${\bf t}_0=\left[0, 0\right]^{\rm T}$ to receive reference position ${\bf r}_0=\left[0, 0\right]^{\rm T}$, in which entry $\sigma_{l_{\rm r},l_{\rm t}}$ denotes the channel coefficient between the $l_{\rm t}$-th transmit path and the $l_{\rm r}$-th receive path with $l_{\rm t} = 1,\cdots, L_{\rm t}$ and $l_{\rm r} = 1,\cdots, L_{\rm r}$. Thus, the channel response between ${\bf t}_{0}$ and ${\bf r}_{0}$ is the linear superposition of the elements in the PRM, i.e., 
\begin{equation}
	h\left({\bf t}_0, {\bf r}_0\right)={\bf 1}^{\rm H}_{L_{\rm r}\times 1}{\bf \Sigma}{\bf 1}_{L_{\rm t}\times 1}.
\end{equation}

Since the moving distance (several wavelengths long) of an MA  is much smaller than the signal propagation distance between the Tx and Rx, the far-field condition can be easily satisfied. Hence, the plane-wave model is adopted. In other words, for different positions of the Tx-/Rx-MA, only the phase of the channel coefficient changes, while the AoD, AoA, and amplitude of the channel coefficient for each channel path are constant. As shown in Fig. \ref{fig:ma_system}, denote $\theta_{\rm t}^{l_{\rm t}}\in\left[-\pi/2, \pi/2\right]$ and $\phi_{\rm t}^{l_{\rm t}}\in\left[-\pi/2, \pi/2\right]$ as the elevation and azimuth AoDs of the $l_{\rm t}$-th transmit path, respectively. According to basic geometry, when the Tx-MA is located at position ${\bf t}=\left[x_{\rm t}, y_{\rm r}\right]^{\rm T}$, the signal propagation distance for the $l_{\rm t}$-th transmit path is changed by $\rho_{\rm t}^{l_{\rm t}}\left(x_{\rm t},y_{\rm t}\right)=x_{\rm t}\cos\theta_{\rm t}^{l_{\rm t}}\sin\phi_{\rm t}^{l_{\rm t}}+y_{\rm t}\sin\theta_{\rm t}^{l_{\rm t}}$ compared to the reference position ${\bf t}_0=\left[0, 0\right]^{\rm T}$. It indicates that there is a phase difference of $2\pi \rho_{\rm t}^{l_{\rm t}}\left(x_{\rm t}, y_{\rm t}\right)/\lambda$ in the channel response of the $l_{\rm t}$-th transmit path with respect to the reference position ${\bf t}_0$, where $\lambda$ is the wavelength. Similarly, for any position ${\bf r}=\left[x_{\rm r}, y_{\rm r}\right]^{\rm T}$ in the receive region, the signal propagation distance is $\rho_{\rm r}^{l_{\rm r}}\left(x_{\rm r},y_{\rm r}\right)=x_{\rm r}\cos\theta_{\rm r}^{l_{\rm r}}\sin\phi_{\rm r}^{l_{\rm r}}+y_{\rm r}\sin\theta_{\rm r}^{l_{\rm r}}$ for the $l_{\rm r}$-th receive path, and the corresponding phase difference is $2\pi \rho_{\rm r}^{l_{\rm r}}\left(x_{\rm r},y_{\rm r}\right)/\lambda$ with respect to the reference position ${\bf r}_0=\left[0, 0\right]^{\rm T}$. Therefore, the channel response between the Tx-MA located at ${\bf t}=\left[x_{\rm t}, y_{\rm t}\right]^{\rm T}$ and Rx-MA located at positions ${\bf r}=\left[x_{\rm r}, y_{\rm r}\right]^{\rm T}$ can be represented as
\begin{equation}\label{channel_response}
	h\left({\bf t, r}\right)={\bf f\left(r\right)}^{\rm H}{\bf \Sigma}{\bf g\left(t\right)},
\end{equation}
where ${\bf g\left(r\right)}\in\mathbb{C}^{L_{\rm t}}\times 1$ and ${\bf f\left(t\right)}\in\mathbb{C}^{L_{\rm r}\times 1}$ denote the transmit and receive field-response vectors (FRVs) to account for phase differences of all channel paths, given by \cite{zhu2023movable}
\begin{equation}\label{frv}
	\left\{
	\begin{aligned}
		{\bf g\left(t\right)}&=\left[e^{j\frac{2\pi}{\lambda}\rho_{\rm t}^{l_{\rm t}}\left(x_{\rm t}, y_{\rm t}\right)}\right]_{1\leq l_{\rm t}\leq L_{\rm t}},\\
		{\bf f\left(r\right)}&=\left[e^{j\frac{2\pi}{\lambda}\rho_{\rm r}^{l_{\rm r}}\left(x_{\rm r}, y_{\rm r}\right)}\right]_{1\leq l_{\rm r}\leq L_{\rm r}}.
	\end{aligned}
	\right.
\end{equation}
It is worth noting that the channel response in \eqref{channel_response} is a function of the positions of the Tx-MA and Rx-MA, indicating that the channel condition can be changed by moving the MAs. Therefore, for MA-aided communication systems, the positions of MAs can be reconfigured to obtain performance gains, e.g., signal power improvement, interference mitigation, and spatial multiplexing\cite{zhu2023mov}. 

Nevertheless, accurate CSI between the Tx and Rx regions is required to achieve such performance gains. Since the Tx and Rx regions are continuous, the number of candidate positions for Tx-MA and Rx-MA, i.e., ${\bf t}$ and ${\bf r}$ in \eqref{channel_response}, is infinite. It is infeasible to move the Tx-/Rx-MA to all candidate positions in the entire spatial region for channel measurements due to the extremely high pilot overhead and time consumption required. Moreover, existing channel estimation methods for FPA systems cannot be directly applied to MA systems. This is because these methods can only estimate the channel responses between the positions where the antennas are located. Hence, a new channel estimation framework for MA systems is required, which can reconstruct the complete CSI between the entire Tx and Rx regions based on a small number of channel measurements at a finite number of MA positions.

\section{Proposed Framework for Channel Estimation}\label{section_3}

In this section, we propose the compressed sensing-based channel estimation framework for MA communication systems. First, we derive a discrete-form channel representation which can approximate the multi-path channel in \eqref{channel_response} by quantizing AoDs/AoAs. Then, based on this representation, a compressed sensing-based channel estimation method is proposed, which can estimate the AoDs, AoAs, and PRM of the MPCs by conducting finite channel measurements at different MA positions. Finally, complete CSI of the entire Tx and Rx regions are reconstructed with the recovered MPC information.

\subsection{Discrete-Form Representation of Multi-Path Channel}


For channel estimation, since the AoDs, AoAs, and PRM of the MPCs are unknown to the transceivers, it is necessary to give a universal representation of the multi-path channel in \eqref{channel_response} with any arbitrary AoDs and AoAs. To this end, a 4-tuple function $\tilde{ \sigma}\left(\varphi_{\rm t}, \vartheta_{\rm t}, \varphi_{\rm r}, \vartheta_{\rm r}\right)$, namely path-response function (PRF), is defined to represent the MPCs in the propagation environment. For notational simplicity, the virtual AoDs are defined as $\varphi_{\rm t}=\cos\theta_{\rm t}\sin\phi_{\rm t}$ and $\vartheta_{\rm t}=\sin\theta_{\rm t}$. Similarly, the virtual AoAs are defined as $\varphi_{\rm r}=\cos\theta_{\rm r}\sin\phi_{\rm r}$ and $\vartheta_{\rm r}=\sin\theta_{\rm r}$. According to the definition, each of the virtual AoDs/AoAs ranges from $-1$ to $1$. For a channel with $L=L_{\rm t}\times L_{\rm r}$ MPCs, the PRF can be expressed as a superposition of $L$ delta impulse functions, i.e., 
\begin{equation}\label{prf}
	\begin{aligned}
		&\tilde{ \sigma}\left(\varphi_{\rm t}, \vartheta_{\rm t}, \varphi_{\rm r}, \vartheta_{\rm r}\right)=\sum_{l_{\rm t}=1}^{L_{\rm t}}\sum_{l_{\rm r}=1}^{L_{\rm r}}\sigma_{l_{\rm t},l_{\rm r}}\times\\
		&\delta\left(\varphi_{\rm t}-\varphi_{\rm t}^{l_{\rm t}},\vartheta_{\rm t}-\vartheta_{\rm t}^{l_{\rm t}},\varphi_{\rm r}-\varphi_{\rm r}^{l_{\rm r}},\vartheta_{\rm r}-\vartheta_{\rm r}^{l_{\rm r}}\right),
	\end{aligned}
\end{equation}
where $\delta\left(x_1, x_2, x_3, x_4\right)$ is defined as the 4-tuple impulse function, i.e., 
\begin{equation}
	\delta\left(x_1,x_2,x_3,x_4\right)=\left\{
	\begin{aligned}
		&+\infty, ~~~x_1=x_2=x_3=x_4=0,\\
		&0, ~~~~~{\rm otherwise},
	\end{aligned}	
	\right.
\end{equation}
and satisfies
\begin{equation}
	\begin{small}
		\begin{aligned}
		&\int_{-\infty}^{+\infty}\int_{-\infty}^{+\infty}\int_{-\infty}^{+\infty}\int_{-\infty}^{+\infty}\delta\left(x_1, x_2, x_3, x_4\right)dx_1dx_2dx_3dx_4 = 1,\\
		&\int_{-\infty}^{+\infty}\int_{-\infty}^{+\infty}\int_{-\infty}^{+\infty}\int_{-\infty}^{+\infty}f\left(x_1, x_2, x_3, x_4\right)\times \delta\left(x_1-x_1^0, x_2-x_2^0,\right.\\
		&\left.x_3-x_3^0, x_4-x_4^0\right)dx_1 dx_2 dx_3 dx_4=f\left(x_1^0, x_2^0, x_3^0, x_4^0\right),
	\end{aligned}
	\end{small}
\end{equation}
where $f\left(x_1,x_2,x_3,x_4\right)$ is an arbitrary 4-tuple function (with certain smoothness).

Moreover, $\varphi_{\rm t}^{l_{\rm t}}=\cos\theta_{\rm t}^{l_{\rm t}}\sin\phi_{\rm t}^{l_{\rm t}}, \vartheta_{\rm t}^{l_{\rm t}}=\sin\theta_{\rm t}^{l_{\rm t}}$ and $\varphi_{\rm r}^{l_{\rm r}}=\cos\theta_{\rm r}^{l_{\rm r}}\sin\phi_{\rm r}^{l_{\rm r}}, \vartheta_{\rm r}^{l_{\rm r}}=\sin\theta_{\rm r}^{l_{\rm r}}$ are the virtual AoDs and AoAs for the $l_{\rm t}$-th transmit path and $l_{\rm r}$-th receive path with $l_{\rm t} = 1,\cdots,L_{\rm t}$ and $l_{\rm r} = 1,\cdots,L_{\rm r}$, respectively. $\sigma_{l_{\rm t},l_{\rm r}}$ is the channel coefficient of the path with virtual AoDs $\vartheta_{\rm t}^{l_{\rm t}}, \varphi_{\rm t}^{l_{\rm t}}$ and AoAs $\vartheta_{\rm r}^{l_{\rm r}}, \varphi_{\rm r}^{l_{\rm r}}$.
As such, the channel response of a multi-path channel in \eqref{channel_response} can be rewritten as
\begin{equation}\label{channel_response_continuous}
	\begin{aligned}
		&h\left({\bf t,r}\right)=\int_{-1}^{1}\int_{-1}^{1}\int_{-1}^{1}\int_{-1}^{1}e^{-j\frac{2\pi}{\lambda}\left(x_{\rm r}\varphi_{\rm r}+y_{\rm r}\vartheta_{\rm r}\right)}\times\\
		&\tilde{ \sigma}\left(\varphi_{\rm t},\vartheta_{\rm t},\varphi_{\rm r},\vartheta_{\rm r}\right)\times e^{j\frac{2\pi}{\lambda}\left(x_{\rm t}\varphi_{\rm t}+y_{\rm t}\vartheta_{\rm t}\right)}d\varphi_{\rm t}d\vartheta_{\rm t}d\varphi_{\rm r}d\vartheta_{\rm r},
	\end{aligned}
\end{equation} 
in which phase terms $e^{j\frac{2\pi}{\lambda}\left(x_{\rm t}\varphi_{\rm t}+y_{\rm t}\vartheta_{\rm t}\right)}$ and $e^{-j\frac{2\pi}{\lambda}\left(x_{\rm r}\varphi_{\rm r}+y_{\rm r}\vartheta_{\rm r}\right)}$ represent the phase differences between ${\bf t}, {\bf r}$ and reference points ${\bf t}_0, {\bf r}_0$, respectively. For an MPC with virtual AoDs $\varphi_{\rm t}^{l_{\rm t}}, \vartheta_{\rm t}^{l_{\rm t}}$ and virtual AoAs $\varphi_{\rm r}^{l_{\rm r}}, \vartheta_{\rm r}^{l_{\rm r}}$, the phase therms become $e^{j\frac{2\pi}{\lambda}\left(x_{\rm t}\varphi_{\rm t}^{l_{\rm t}}+y_{\rm t}\vartheta_{\rm t}^{l_{\rm t}}\right)}$ and $e^{-j\frac{2\pi}{\lambda}\left(x_{\rm r}\varphi_{\rm r}^{l_{\rm r}}+y_{\rm r}\vartheta_{\rm r}^{l_{\rm r}}\right)}$, which are the $l_{\rm t}$-th entry in ${\bf g}\left({\bf t}\right)$ and the $l_{\rm r}$-th entry in ${\bf f}\left({\bf r}\right)^{\rm H}$, respectively.

Note that in practical systems for channel estimation, the virtual AoDs and AoAs in \eqref{channel_response_continuous} can be any value within the range from $-1$ to $1$, which contains infinite real numbers. The virtual AoDs and AoAs cannot be perfectly recovered by a limited number of channel measurements in the confined Tx and Rx regions. In this regard, a discrete-form approximation of the channel response is required to facilitate the estimation of MPCs. Thus, we uniformly quantize the virtual AoDs and AoAs (ranging from $-1$ to $1$) into $N$ grids, i.e.,
\begin{equation}\label{dis_cosangle_tx}
	\left\{
		\begin{aligned}
			\tilde{\varPhi}_{\rm t}=&\left\{\left.\tilde{\varphi}_{\rm t}^{n_{{\rm t}x}} = -1+\frac{2n_{{\rm t}x}-1}{N}\right|1\leq n_{{\rm t}x}\leq N\right\}, \\
			\tilde{\varTheta}_{\rm t}=& \left\{\left.\tilde{\vartheta}_{\rm t}^{n_{{\rm t}y}}=-1+\frac{2n_{{\rm t}y}-1}{N}\right|1\leq n_{{\rm t}y}\leq N\right\},\\
			\tilde{\varPhi}_{\rm r}=&\left\{\left.\tilde{\varphi}_{\rm r}^{n_{{\rm r}x}}=-1+\frac{2n_{{\rm r}x}-1}{N}\right|1\leq n_{{\rm r}x}\leq N\right\}, \\
			\tilde{\varTheta}_{\rm r}=& \left\{ \left.\tilde{\vartheta}_{\rm r}^{n_{{\rm r}y}}=-1+\frac{2n_{{\rm r}y}-1}{N}\right|1\leq n_{{\rm r}y}\leq N\right\}. \\		
		\end{aligned}
	\right.
\end{equation}
Note that in \eqref{dis_cosangle_tx} the angular resolution is $2/N$, indicating that more accurate virtual angles can be estimated with a larger $N$. Then, the channel response in \eqref{channel_response_continuous} can be approximated as the following discrete form:
\begin{equation}\label{channel_response_discrete}
	\begin{aligned}
		&h_{\rm d}\left({\bf t,r}\right)=
		\sum_{n_{{\rm t}x}=1}^{N}\sum_{n_{{\rm t}y}=1}^{N}\sum_{n_{{\rm r}x}=1}^{N}\sum_{n_{{\rm r}y}=1}^{N}e^{j\frac{2\pi}{\lambda}\left(x_{\rm t}\tilde{\varphi}_{\rm t}^{n_{{\rm t}x}}+y_{\rm t}\tilde{\vartheta}_{\rm t}^{n_{{\rm t}y}}\right)}\times\\ 
		& \tilde{ \sigma}_{\rm d}\left(n_{{\rm t}x},n_{{\rm t}y},n_{{\rm r}x},n_{{\rm r}y}\right)\times e^{-j\frac{2\pi}{\lambda}\left(x_{\rm r}\tilde{\varphi}_{\rm r}^{n_{{\rm r}x}}+y_{\rm r}\tilde{\vartheta}_{\rm r}^{n_{{\rm r}y}}\right)}.
	\end{aligned}
\end{equation}
$\tilde{ \sigma}_{\rm d}\left(n_{{\rm t}x},n_{{\rm t}y},n_{{\rm r}x},n_{{\rm r}y}\right)$ is the discrete PRF and can be represented as
\begin{equation}\label{discrete_prf}
	\begin{aligned}
		&\tilde{ \sigma}_{\rm d}\left(n_{{\rm t}x}, n_{{\rm t}y}, n_{{\rm r}x}, n_{{\rm r}y}\right)=\sum_{l_{\rm t}=1}^{L_{\rm t}}\sum_{l_{\rm r}=1}^{L_{\rm r}}\sigma_{l_{\rm t},l_{\rm r}}\times\\
		&\delta_{\rm d}\left[\tilde{\varphi}_{\rm t}^{n_{{\rm t}x}}-\tilde{\varphi}_{\rm t}^{l_{\rm t}},\tilde{\vartheta}_{\rm t}^{n_{{\rm t}y}}-\tilde{\vartheta}_{\rm t}^{l_{\rm t}},\tilde{\varphi}_{\rm r}^{n_{{\rm r}x}}-\tilde{\varphi}_{\rm r}^{l_{\rm r}},\tilde{\vartheta}_{\rm r}^{n_{{\rm r}y}}-\tilde{\vartheta}_{\rm r}^{l_{\rm r}}\right],
	\end{aligned}
\end{equation}
where $\delta_{\rm d}\left[x_1,x_2,x_3,x_4\right]$ represents the 4-tuple discrete impulse function, i.e.,
\begin{equation}
	\delta_{\rm d}\left[x_1,x_2,x_3,x_4\right]=\left\{
	\begin{aligned}
		&1, ~~~x_1=x_2=x_3=x_4=0,\\
		&0, ~~~~~{\rm otherwise}.
	\end{aligned}	
	\right.
\end{equation}
$\tilde{\varphi}_{\rm t}^{l_{\rm t}}, \tilde{\vartheta}_{\rm t}^{l_{\rm t}}$ and $\tilde{\varphi}_{\rm r}^{l_{\rm r}}, \tilde{\vartheta}_{\rm r}^{l_{\rm r}}$ are quantized virtual AoDs and AoAs for the path with virtual AoDs $\varphi_{\rm t}^{l_{\rm t}}, \vartheta_{\rm t}^{l_{\rm t}}$ and AoAs $\varphi_{\rm r}^{l_{\rm r}}, \vartheta_{\rm r}^{l_{\rm r}}$, respectively, i.e., 
\begin{equation}
	\begin{aligned}
		\tilde{\varphi}_{\rm t}^{l_{\rm t}}=\mathop{\arg\min}\limits_{\tilde{\varphi}_{\rm t}^{n_{{\rm t}x}}\in \tilde{\varPhi}_{\rm t}} \left|\varphi_{\rm t}^{l_{\rm t}}-\tilde{\varphi}_{\rm t}^{n_{{\rm t}x}}\right|,~
		\tilde{\vartheta}_{\rm t}^{l_{\rm t}}=\mathop{\arg\min}\limits_{\tilde{\vartheta}_{\rm t}^{n_{{\rm t}y}}\in \tilde{\varTheta}_{\rm t}} \left|\vartheta_{\rm t}^{l_{\rm t}}-\tilde{\vartheta}_{\rm t}^{n_{{\rm t}y}}\right|,\\
		\tilde{\varphi}_{\rm r}^{l_{\rm r}}=\mathop{\arg\min}\limits_{\tilde{\varphi}_{\rm r}^{n_{{\rm r}x}}\in \tilde{\varPhi}_{\rm r}} \left|\varphi_{\rm r}^{l_{\rm r}}-\tilde{\varphi}_{\rm r}^{n_{{\rm r}x}}\right|,~
		\tilde{\vartheta}_{\rm r}^{l_{\rm r}}=\mathop{\arg\min}\limits_{\tilde{\vartheta}_{\rm r}^{n_{{\rm r}y}}\in \tilde{\varTheta}_{\rm r}} \left|\vartheta_{\rm r}^{l_{\rm r}}-\tilde{\vartheta}_{\rm r}^{n_{{\rm r}y}}\right|.\\
	\end{aligned}
\end{equation}
Accordingly, an error exists between the quantized virtual angles $\tilde{\varphi}_{\rm t}^{l_{\rm t}}, \tilde{\vartheta}_{\rm t}^{l_{\rm t}}, \tilde{\varphi}_{\rm r}^{l_{\rm r}}, \tilde{\vartheta}_{\rm r}^{l_{\rm r}}$ and the actual virtual angles $\varphi_{\rm t}^{l_{\rm t}}, \vartheta_{\rm t}^{l_{\rm t}}, \varphi_{\rm r}^{l_{\rm r}}, \vartheta_{\rm r}^{l_{\rm r}}$, leading to a mismatch between the discrete channel response in \eqref{channel_response_discrete} and the actual channel response in \eqref{channel_response_continuous}. In this regard, quantization error $e\left({\bf t,r}\right)$ is defined to measure the mismatch, i.e., 
\begin{equation}
	e\left({\bf t,r}\right)=h\left({\bf t,r}\right) - h_{\rm d}\left({\bf t,r}\right).
\end{equation}

Rewriting the discrete channel response in \eqref{channel_response_discrete} in a matrix form, the discrete channel model can be obtained as
\begin{equation}\label{channel_response_matrix}
	h\left({\bf t,r}\right) = \tilde{{\bf f}}\left({\bf r}\right)^{\rm H}\tilde{{\bf \Sigma}}\tilde{{\bf g}}\left({\bf t}\right)+e\left({\bf t,r}\right),
\end{equation}
in which the $n_{\rm r}$-th row and $n_{\rm t}$-th column entry in $\tilde{{\bf \Sigma}}\in\mathbb{C}^{N^2\times N^2}$ is given by
\begin{equation}\label{temp_1}
		\left\{
	\begin{aligned}
		\left[\tilde{{\bf \Sigma}}\right]_{n_{\rm r},n_{\rm t}}=&\tilde{ \sigma}_{\rm d}\left(n_{{\rm t}x},n_{{\rm t}y},n_{{\rm r}x}, n_{{\rm r}y}\right),\\
		n_{{\rm t}x} =& n_{\rm t} \bmod N,~ n_{{\rm t}y} = \biggl\lfloor \frac{n_{\rm t}}{N}\biggr\rfloor +1,\\
		n_{{\rm r}x} =& n_{\rm r} \bmod N,~~ n_{{\rm r}y} = \biggl\lfloor \frac{n_{\rm r}}{N}\biggr\rfloor +1.
	\end{aligned}
	\right.
\end{equation}
Note that the above matrix $\tilde{{\bf \Sigma}}$ contains all the quantized virtual AoDs, quantized virtual AoAs, and their corresponding complex coefficients, so we call it the {\em discrete PRM} of the MA. In addition, $\tilde{{\bf g}}\left({\bf t}\right)\in\mathbb{C}^{N^2\times 1}$ and $\tilde{{\bf f}}\left({\bf r}\right)\in\mathbb{C}^{N^2\times 1}$ are the discrete FRVs and can be represented as
\begin{equation}\label{discrete_frv}
	\left\{
	\begin{aligned}
		\tilde{{\bf g}}\left({\bf t}\right)&=\left[e^{j\frac{2\pi}{\lambda}y_{\rm t}\tilde{\vartheta}_{\rm t}^{n_{{\rm t}y}}}\right]_{1\leq n_{{\rm t}y}\leq N}\otimes \left[e^{j\frac{2\pi}{\lambda}x_{\rm t}\tilde{\varphi}_{\rm t}^{n_{{\rm t}x}}}\right]_{1\leq n_{{\rm t}x}\leq N},\\
		\tilde{{\bf f}}\left({\bf r}\right)&=\left[e^{j\frac{2\pi}{\lambda}y_{\rm r}\tilde{\vartheta}_{\rm r}^{n_{{\rm r}y}}}\right]_{1\leq n_{{\rm r}y}\leq N}\otimes \left[e^{j\frac{2\pi}{\lambda}x_{\rm r}\tilde{\varphi}_{\rm r}^{n_{{\rm r}x}}}\right]_{1\leq n_{{\rm r}x}\leq N}.
	\end{aligned}
	\right.
\end{equation}
Substituting \eqref{channel_response_matrix} into \eqref{signal_model}, the received signal can be represented as
\begin{equation}\label{channel_model_discrete}
	v\left({\bf t, r}\right)=\left({\bf {\tilde{f}}\left(r\right)}^{\rm H}{\bf {\tilde{\Sigma}} {\tilde{g}}\left(t\right)}+e\left({\bf t, r}\right)\right)\sqrt{p_{\rm t}}s+z,
\end{equation}

\subsection{Channel Estimation}

The discrete PRM in \eqref{temp_1} involves the path responses among all the quantized angles, which can be regarded as an `over-sampling' in the angular domain. Thus, the discrete PRM $\tilde{{\bf \Sigma}}$ is  a sparse matrix with only a small number of non-zero entries, making it possible to utilize compressed sensing for recovering the discrete PRM. Define path response vector ${\bf u} = {\rm vec}\left(\tilde{{\bf \Sigma}}\right)\in\mathbb{C}^{N^4\times 1}$ as the vectorized discrete PRM, and then we have
\begin{equation}
	v\left({\bf t, r}\right)=\left[\left(\tilde{\bf g}\left({\bf t}\right)^{\rm T}\otimes \tilde{\bf f}\left({\bf r}\right)^{\rm H}\right){\rm vec}\left(\tilde{{\bf \Sigma}}\right)+e\left({\bf t,r}\right)\right]\sqrt{p_{\rm t}}s+z.
\end{equation}

For MA systems, the Tx-MA and Rx-MA can move in the Tx and Rx regions, indicating that the positions of Tx-MA and Rx-MA can be changed over time for acquiring the MPC information. Specifically, in the $m$-th time slot, the Tx-MA and Rx-MA move to positions ${\bf t}^{m}=\left[x_{\rm t}^{m}, y_{\rm t}^{m}\right]^{\rm T}$ and ${\bf r}^{m}=\left[x_{\rm r}^{m}, y_{\rm r}^{m}\right]^{\rm T}$ for channel measurement. Without loss of generality, let the transmit pilot be $s=1$ during the channel estimation stage. Then, collecting all the $M$ received signals $v^{m}\left({\bf t}^{m},{\bf r}^{m}\right)$ from the transmitted pilots called the channel measurements and stacking them into vector ${\bf v}\in\mathbb{C}^{M\times 1}$, we have
\begin{equation}\label{signal_model_2}
	\begin{aligned}
		{\bf v}=&\left\{\begin{bmatrix}
			\tilde{\bf g}\left({\bf t}^1\right)^{\rm T}\otimes \tilde{\bf f}\left({\bf r}^1\right)^{\rm H}\\
			\vdots \\
			\tilde{\bf g}\left({\bf t}^M\right)^{\rm T}\otimes \tilde{\bf f}\left({\bf r}^M\right)^{\rm H}
		\end{bmatrix}
		 {\bf u}+{\bf e}\right\}\sqrt{p_{\rm t}}+{\bf z}\\
		\triangleq& \sqrt{p_{\rm t}}{\bf \Psi}{\bf u}+\sqrt{p_{\rm t}}{\bf e}+{\bf z},
	\end{aligned}
\end{equation}
where ${\bf \Psi}\in \mathbb{C}^{M\times N^4}$ is the measurement matrix. ${\bf e}\in\mathbb{C}^{M\times 1}$ and ${\bf z}\in\mathbb{C}^{M\times 1}$ represent the quantization error vector and the noise vector over $M$ channel measurements, respectively. Then, the MPC information can be obtained by solving the following sparse signal recovery problem
\begin{equation}\label{cs_channel_estimation}
	\begin{aligned}
		&\mathop{\min}\limits_{{\bf u}} ~\Vert {\bf u} \Vert_0,\\
		&{\rm s.t.}~~\Vert {\bf v}-\sqrt{p_{\rm t}}{\bf \Psi}{\bf u} \Vert_2 \leq \Vert{\bf v}\Vert_2\epsilon_0,
	\end{aligned}
\end{equation}
where $\epsilon_0$ is a small positive parameter to guarantee the minimization of the channel estimation error considering the impact of angular quantization and noise. According to compressed sensing theory, despite the high dimensionality of ${\bf u}$, it can be recovered by a few measurements due to its sparsity, i.e., $L_{\rm t}\times L_{\rm r}\ll N^4$, which can significantly reduce the pilot overhead. 

In this paper, the sparse signal recovery problem \eqref{cs_channel_estimation} is solved by the classical OMP algorithm, which can jointly estimate the virtual AoDs, virtual AoAs, and channel coefficients\cite{7458188, li2017millimeter,gao2015spatially}. The details of the OMP-based channel estimation algorithm are summarized in Algorithm \ref{alg:omp}. Specifically, in lines 4 and 5, we recover set $\mathcal{A}$, which contains the virtual AoDs and AoAs of the recovered channel path. Then, in lines 6 and 7, the corresponding complex coefficients are recovered by employing the least-square (LS) estimate \cite{7458188}. In lines 8 and 9, we update the residual, and the OMP algorithm terminates when the normalized residual power is below threshold parameter $\epsilon_0$ in \eqref{cs_channel_estimation}. Then, in line 12, we extract the path response information, i.e., the virtual AoDs, virtual AoAs, and channel coefficients, through ${\bf u}$ and set $\mathcal{A}$. Finally, in line 13, we reconstruct the channel response between the Tx and Rx regions according to \eqref{channel_response_discrete}. Next, we analyze the computational complexity of the channel estimation algorithm. Specifically, in line 4, the complexities of calculating $\left[{\bf\Psi}\right]^{\rm H}_{:,i}{\bf r}^{\left(k\right)}$ and finding the index of the maximum value $j$ are $\mathcal{O}\left(MN^4\right)$ and $\mathcal{O}\left(N^4\right)$, respectively. In line 6, the complexity of calculating ${\bf q}^{\left(k\right)}$ is no larger than $\mathcal{O}\left(MK_{\rm max}^{2}\right)$, where $K_{\rm max}$ denotes the maximum number of iterations. In lines 8 and 9, the complexities of updating ${\bf r}^{\left(k\right)}$ and $\epsilon^{\left(k\right)}$ are $\mathcal{O}\left(MK_{\rm max}\right)$ and $\mathcal{O}\left(M\right)$, respectively. Thus, the total complexity of Algorithm \ref{alg:omp} is $\mathcal{O}\left(K_{\rm max}M\left(N^4+K_{\rm max}^2\right)\right)$.

\begin{small}
	\begin{algorithm}[t] 
	\label{alg:omp}
	\caption{OMP Algorithm.}
	\begin{algorithmic}[1]
		\REQUIRE ${\bf \Psi}, {\bf v}, \epsilon, {\bf t}, {\bf r}$.
		\ENSURE Channel response $h\left({\bf t, r}\right)$. \\
		\STATE Initialization: ${\bf r}^{\left(1\right)} = {\bf v}, \mathcal{A}^{\left(0\right)} = \emptyset, {\bf u}^{\left(0\right)}={\bf 0}, \epsilon^{\left(1\right)}=1$.	
		\STATE Set the iteration index as $k=1$.	
		\WHILE{$\epsilon^{\left(k\right)}\geq \epsilon_0$}
		\STATE Find index: $j= \mathop{\arg\max}\limits_{i \notin \mathcal{A}^{\left(k-1\right)}} \left|\left[{\bf \Psi}\right]_{:,i}^{\rm H}{\bf r}^{\left(k\right)} \right|$.
		\STATE Update $\mathcal{A}^{\left(k\right)}=\mathcal{A}^{\left(k-1\right)}\cup \left\{j\right\}$.
		\STATE Recover channel coefficient ${{\bf q}}^{\left(k\right)}=\frac{1}{\sqrt{p_{\rm t}}}\left[\left(\left[{\bf \Psi}\right]_{:,\mathcal{A}^{\left(k\right)}}\right)^{\rm H}\left[{\bf \Psi}\right]_{:,\mathcal{A}^{\left(k\right)}}\right]^{-1}\left(\left[{\bf \Psi}\right]_{:,\mathcal{A}^{\left(k\right)}}\right)^{\rm H}{\bf v}$.
		\STATE Update $\left[{\bf u}^{\left(k\right)}\right]_{\mathcal{A}^{\left(k\right)}} = {\bf q}^{\left(k\right)}$.
		\STATE Update ${\bf r}^{\left(k\right)} = {\bf v} - \sqrt{p_{\rm t}}\left[{\bf \Psi}\right]_{:,\mathcal{A}^{\left(k\right)}}{{\bf q}}^{\left(k\right)}$.
		\STATE Update $\epsilon^{\left(k\right)}=\frac{\Vert {\bf r}^{\left(k\right)}\Vert_2}{\Vert{\bf v}\Vert_2}$.
		\STATE $k\leftarrow k+1$.
		\ENDWHILE
		\STATE Obtain the AoDs, AoAs, and complex coefficients of the recovered paths using ${\bf u}$ and $\mathcal{A}$.
		\STATE Obtain channel response $h\left({\bf t,r}\right)$ according to \eqref{channel_response_discrete}.
	\end{algorithmic}
\end{algorithm}
\end{small}

\section{Design of Measurement Matrix}\label{section_4}

The design of the measurement matrix is a core issue in compressed sensing theory \cite{6189388,foucart2013invitation}, which may significantly influence the performance of channel estimation for the considered MA communication systems. However, the design of the measurement matrix is challenging in MA systems due to the limited DoFs. Specifically, according to \eqref{signal_model_2}, the measurement matrix can only be modified by changing the MA measurement positions. This is different from channel estimation in conventional MIMO or massive MIMO systems which can flexibly optimize the pilot signals of multiple antennas for designing the measurement matrix \cite{7458188,gao2015spatially}. Hence, in this section, we first analyze the impact of MA measurement positions on the form of the measurement matrix as well as its mutual coherence property. Then, based on the analysis, we propose five setups for MA measurement with deterministic or random positions that can be used for channel estimation.

For compressed sensing, the design of measurement matrix can be mainly divided into two categories, i.e., deterministic matrices and random matrices. Each category utilizes distinct strategies to satisfy the restricted isometry property (RIP), which is a quintessential condition underpinning the success of compressed sensing. Deterministic matrices satisfy the RIP based on their inherent structural characteristics, e.g., chirp sensing codes matrix \cite{applebaum2009chirp}, Toeplitz-structured matrix \cite{4301266}, and binary matrix\cite{7446367}. For random matrices, Gaussian random matrix and Bernoulli random matrix, whose entries obey Gaussian and Bernoulli distribution, respectively, are statistically poised to fulfill RIP with high probabilities \cite{4016283,1542412,WOS:000260684900002}. However, for the considered MA channel estimation problem \eqref{signal_model_2}, since the measurement matrix can only be configured by the MA measurement positions, the DoFs in the measurement matrix design are limited. Thus, conventional measurement matrices cannot be implemented in our case. Moreover, it is challenging for the measurement matrix in \eqref{signal_model_2} to satisfy RIP due to the limited DoFs. Instead of RIP, mutual coherence can be used as an alternative measure of the measurement matrix. Specifically, the cross correlation of two arbitrary columns in the measurement matrix requires to be as low as possible to enhance the recovery performance of \eqref{cs_channel_estimation} \cite{foucart2013invitation}. In other words, the mutual coherence matrix is required to be as close as possible to an identity matrix. In this regard, we consider the mutual coherence as a performance metric for designing the measurement matrix.

\subsection{Mutual Coherence}

According to \eqref{signal_model_2}, the $m$-th row and $n$-th column entry of the measurement matrix is given by
\begin{equation}\label{cs_element}
		\begin{aligned}
			\left[{\bf \Psi}\right]_{m,n}= &e^{j\frac{2\pi}{\lambda}\left(x_{\rm t}^{m}\tilde{\varphi}_{\rm t}^{n_{{\rm t}x}}+y_{\rm t}^{m}\tilde{\vartheta}_{\rm t}^{n_{{\rm t}y}}\right)}\times e^{-j\frac{2\pi}{\lambda}\left(x_{\rm r}^{m}\tilde{\varphi}_{\rm r}^{n_{{\rm r}x}}+y_{\rm r}^{m}\tilde{\vartheta}_{\rm r}^{n_{{\rm r}y}}\right)},
		\end{aligned}
\end{equation}
where $n_{{\rm t}x}, n_{{\rm t}y}, n_{{\rm r}x}$, and $n_{{\rm r}y}$ satisfy
\begin{equation}
	n = N^2\left[N\left(n_{{\rm t}y}-1\right)+n_{{\rm t}x}-1\right]+N\left(n_{{\rm r}y}-1\right)+n_{{\rm r}x},
\end{equation}
and they uniquely determine a pair of virtual AoDs and AoAs according to \eqref{dis_cosangle_tx}.

The mutual coherence is defined as ${\bf C} = \frac{1}{M}{\bf \Psi}^{\rm H}{\bf \Psi}\in\mathbb{C}^{N^4\times N^4}$. The entry in the $n$-th row and $n^{\prime}$-th column in ${\bf C}$, representing the cross correlation of the two columns in measurement matrix ${\bf \Psi}$, can be written as
\begin{equation}\label{correlation}
		\begin{aligned}
		\left[{\bf C}\right]_{n,n^{\prime}} =&\frac{1}{M} \sum_{m=1}^{M}e^{j\frac{2\pi}{\lambda}\left[x_{\rm t}^{m}\left(\tilde{\varphi}_{\rm t}^{n_{{\rm t}x}^{\prime}}-\tilde{\varphi}_{\rm t}^{n_{{\rm t}x}}\right)+y_{\rm t}^{m}\left(\tilde{\vartheta}_{\rm t}^{n_{{\rm t}y}^{\prime}}-\tilde{\vartheta}_{\rm t}^{n_{{\rm t}y}}\right)\right]} \times\\
		& e^{-j\frac{2\pi}{\lambda}\left[x_{\rm r}^{m}\left(\tilde{\varphi}_{\rm r}^{n_{{\rm r}x}^{\prime}}-\tilde{\varphi}_{\rm r}^{n_{{\rm r}x}}\right)+y_{\rm r}^{m}\left(\tilde{\vartheta}_{\rm r}^{n_{{\rm r}y}^{\prime}}-\tilde{\vartheta}_{\rm r}^{n_{{\rm r}y}}\right)\right]}.\\
	\end{aligned}
\end{equation}
To guarantee successful MPC recovery, mutual coherence ${\bf C}$ is desired to be as close as possible to the identity matrix, indicating that different columns in ${\bf \Psi}$ become less correlated \cite{8079818}. A necessary condition is to reduce the cross correlations among $N$ particular columns in ${\bf \Psi}$ that satisfy $n_{{\rm t}y}^{\prime}=n_{{\rm t}y}, n_{{\rm r}x}^{\prime}=n_{{\rm r}x}$ and $n_{{\rm r}y}^{\prime}=n_{{\rm r}y}$ 
(i.e., $N$ particular columns with indices $n = n_{{\rm t}x} = 1,\cdots,N$), because they are likely to have a high coherence with each other\footnote{Without loss of generality, the following analysis takes the recovery of $\tilde{\varphi}_{\rm t}^{l_{\rm t}}, l_{\rm t} = 1,\cdots, L_{\rm t}$, as an example, which is also applicable to the recovery of $\tilde{\vartheta}^{l_{\rm t}}_{\rm t}, l_{\rm t} = 1,\cdots, L_{\rm t}$, and $\tilde{\varphi}^{l_{\rm r}}_{\rm r}, \tilde{\vartheta}^{l_{\rm r}}_{\rm r}, l_{\rm r} = 1,\cdots, L_{\rm r}$.}. Then, for the considered $N$ columns, an effective mutual coherence $\tilde{{\bf C}}\in\mathbb{C}^{N\times N}$ can be defined as
\begin{equation}\label{correlation_2}
	\left[\tilde{{\bf C}}\right]_{n_{{\rm t}x},n_{{\rm t}x}^{\prime}} =\frac{1}{M} \sum_{m=1}^{M}e^{j\frac{2\pi}{\lambda}x_{\rm t}^{m}\left(\tilde{\varphi}^{n^{\prime}_{{\rm t}x}}_{\rm t}-\tilde{\varphi}^{n_{{\rm t}x}}_{\rm t}\right)}.
\end{equation}
To reduce the mutual coherence in \eqref{correlation_2}, we introduce an effective measurement matrix $\tilde{{\bf \Psi}}\in\mathbb{C}^{M\times N}$ with entry in the $m$-th row and $n_{{\rm t}x}$-th column given by
\begin{equation}\label{cs_element_2}
	\left[\tilde{{\bf \Psi}}\right]_{m,n_{{\rm t}x}}= e^{j\frac{2\pi}{\lambda}x_{\rm t}^{m}\tilde{\varphi}^{n_{{\rm t}x}}_{\rm t}}.
\end{equation}
Effective mutual coherence $\tilde{{\bf C}}=\frac{1}{M}{ \tilde{\bf\Psi}}^{\rm H}{ \tilde{\bf\Psi}}$ can be optimized by designing effective measurement matrix $\tilde{{\bf \Psi}}$. Notably, when $y_{\rm t}^m = x_{\rm r}^m = y_{\rm r}^m = 0, m = 1,\cdots,M$, i.e., the Tx-MA moves along the $x$-axis while the Rx-MA is fixed at the reference point, the following relationship between ${\bf \Psi}$ and $\tilde{{\bf \Psi}}$ holds
\begin{equation}\label{effective_measurement_matrix}
	{\bf \Psi} = {\bf 1}_{1\times N}\otimes \tilde{{\bf \Psi}} \otimes {\bf 1}_{1\times N^2}.
\end{equation}
Then, substituting \eqref{effective_measurement_matrix} into \eqref{signal_model_2}, we have
\begin{equation}\label{signal_model_3}
	\begin{aligned}
		{\bf v} = &\sqrt{p_{\rm t}}\left({\bf 1}_{1\times N}\otimes{\tilde{{\bf \Psi}}}\otimes{\bf 1}_{1\times N^2}\right){\bf u}+\sqrt{p_{\rm t}}{\bf e}+{\bf z}\\
		=&\sqrt{p_{\rm t}}\tilde{{\bf \Psi}}\tilde{{\bf u}}+\sqrt{p_{\rm t}}{\bf e}+{\bf z},
	\end{aligned}
\end{equation}
where $\tilde{{\bf u}}\in\mathbb{C}^{N\times 1}$ is defined as the simplified path response vector with the $n_{{\rm t}x}$-th entry given by
\begin{equation}
	\begin{aligned}
		\left[\tilde{{\bf u}}\right]_{n_{{\rm t}x}}
		= \sum_{q = 1}^{N^2}\sum_{p=1}^{N}\left[{\bf u}\right]_{\left(p-1\right)N^3+\left(n_{{\rm t}x}-1\right)N^2+q}.
	\end{aligned}
\end{equation}
This indicates that $\tilde{{\bf u}}$ contains all virtual AoDs $\tilde{\varphi}_{\rm t}^{n_{{\rm t}x}}, n_{{\rm t}x} = 1,\cdots, N$, and the $n_{{\rm t}x}$-th entry in $\tilde{{\bf u}}$ is the summation of all the path responses from virtual AoD $\tilde{\varphi}_{\rm t}^{n_{{\rm t}x}}$.

It can be observed from \eqref{signal_model_3} that the left multiplication effective measurement matrix $\tilde{{\bf \Psi}}$ can be regarded as performing a transform on $\tilde{{\bf u}}$, which is similar to the Fourier transform. Specifically, the angular domain that contains virtual angles with path response can be regarded as the frequency domain, while the locational domain with channel response can be regarded as the time domain. This alignment permits leveraging Fourier properties to design effective measurement matrix $\tilde{{\bf \Psi}}$, thereby reducing mutual coherence in \eqref{correlation_2}. Then, transform $\mathcal{F}_{\tilde{{\bf \Psi}}}\left(\cdot\right)$ can be defined as 
\begin{equation}\label{dtft}
		\tilde{v}\left(x_{\rm t}\right) = \mathcal{F}_{\tilde{{\bf \Psi}}}\left(\tilde{{\bf u}}\right) = \sum_{n_{{\rm t}x}=1}^{N}\left[\tilde{{\bf u}}\right]_{n_{{\rm t}x}} e^{j\frac{2\pi}{\lambda}x_{\rm t}\left(-1+\frac{2n_{{\rm t}x}-1}{N}\right)},
\end{equation}
which transforms $\tilde{{\bf u}}$ from the angular domain to the locational domain, and function $\tilde{v}\left(x_{\rm t}\right)$ represents the channel response along the $x$-axis in the Tx region, which can be thought as the Fourier series of the sequence $\left[0,\cdots,0,\tilde{{\bf u}}^{\rm T},0,\cdots,0\right]$. 

As $N\rightarrow+\infty$ (i.e., containing all the angular domain information), the path response vector $\tilde{{\bf u}}$ (multiplied by $\frac{N}{2}$) converges to a continuous angular domain function $\tilde{u}\left(\varphi_{\rm t}\right)$ with complex values in interval $\left[-1,1\right]$, representing the complex coefficient of a path with a certain virtual AoD, and zero value elsewhere, indicating that the path response in the angular domain is bandlimited. Then, the transform $\mathcal{F}_{\bf \tilde{\Psi}}\left(\cdot\right)$ becomes a continuous transform $\mathcal{F}_{\bf \tilde{\Psi}}^{\rm c}\left(\cdot\right)$, i.e.,
\begin{equation}\label{con_dtft}
	\begin{small}
		\begin{aligned}
		&\tilde{v}\left(x_{\rm t}\right) =\lim_{N\rightarrow+\infty} \sum_{n_{{\rm t}x}=1}^{N}\left[\tilde{{\bf u}}\right]_{n_{{\rm t}x}}e^{j\frac{2\pi}{\lambda}x_{\rm t}\left(-1+\frac{2n_{{\rm t}x}-1}{N}\right)} \\
		&= \lim_{N\rightarrow+\infty}\sum_{n_{{\rm t}x}=1}^{N}\frac{N}{2}\left[\tilde{{\bf u}}\right]_{n_{{\rm t}x}}e^{j\frac{2\pi}{\lambda}x_{\rm t}\left(-1+\frac{2n_{{\rm t}x}-1}{N}\right)}\times \frac{2}{N}\\
		&=\int_{-1}^{1}\tilde{u}\left(\varphi_{\rm t}\right)e^{j\frac{2\pi}{\lambda}x_{\rm t}\varphi_{\rm t}}d\varphi_{\rm t}\triangleq\mathcal{F}_{\bf \tilde{\Psi}}^{\rm c}\left[\tilde{u}\left(\varphi_{\rm t}\right)\right].
	\end{aligned}
	\end{small}
\end{equation}
Thus, the inverse transform ${\mathcal{F}_{\bf \tilde{\Psi}}^{\rm c}}^{-1}\left(\cdot\right)$ can be expressed as 
\begin{equation}\label{con_idtft}
	\begin{small}
		\begin{aligned}
			\tilde{u}\left(\varphi_{\rm t}\right) = {\mathcal{F}^{\rm c}_{\tilde{{\bf \Psi}}}}^{-1}\left[\tilde{v}\left(x_{\rm t}\right)\right]=\lim_{N\rightarrow+\infty}\frac{1}{\lambda}\int_{-\frac{N\lambda}{4}}^{\frac{N\lambda}{4}}\tilde{v}\left(x_{\rm t}\right)e^{-j\frac{2\pi}{\lambda}x_{\rm t}\varphi_{\rm t}}dx_{\rm t},
		\end{aligned}
	\end{small}
\end{equation}
which transforms the channel responses in the locational domain back to the path responses in the angular domain. 

Note that \eqref{con_idtft} indicates that the path response with arbitrarily high angular resolution can be recovered by performing continuous channel measurements in an infinite region, i.e., $\frac{N\lambda}{4}\rightarrow+\infty$. However, in practical MA communication systems, channel measurements can only be performed at discrete positions $\left\{x_{\rm t}^{m}\right\}, m = 1,\cdots, M$, in a confined region $r$, i.e., $-\frac{r\lambda}{2}\leq x_{\rm t}^{m}\leq\frac{r\lambda}{2}$ and $r\leq R$, which can be regarded as sampling in the locational domain. Thus, in the following, we will separately analyze the impact of sampling and finite region for channel measurements on the performance of path response recovery.

As aforementioned, $\tilde{u}\left(\varphi_{\rm t}\right)$ is bandlimited, indicating that only a discrete sequence of samples of $\tilde{v}\left(x_{\rm t}\right)$ (rather than the whole continuous $\tilde{v}\left(x_{\rm t}\right)$) is required to guarantee the full recovery of the path responses according to the Shannon sampling theorem. Since there is no prior information of $\tilde{u}\left(\varphi_{\rm t}\right)$, the equally spaced placement of MA measurement positions, i.e., uniform sampling in the locational domain, should be optimal. For uniform sampling, the measurement positions can be represented as $x_{\rm t}^{m} = -\frac{r\lambda}{2}+\left(m-1\right)\Delta x, m = 1,\cdots,M$, where $\Delta x$ represents the sampling interval length, i.e., the spacing between adjacent MA measurement positions. Note that $\left(M-1\right)\Delta x = r\lambda$ holds for the number $M$ of measurement positions and channel measurement region $r$. The sampled channel response can be represented as $\tilde{v}_{\rm s}\left(x_{\rm t}\right)= \tilde{v}\left(x_{\rm t}^{m}\right)\times\frac{N\Delta x}{2r}\sum_{m=1}^{M}\delta\left(x_{\rm t}-x_{\rm t}^{m}\right)$. Then, we have the following.
\begin{criterion}\label{uniform_sampling}
	For uniform sampling at positions $\left\{x_{\rm t}^{m}\right\}$ in the locational domain with an infinite channel measurement region, i.e., $r\rightarrow+\infty$ and $M\rightarrow +\infty$, the sampling interval length $\Delta x$ should be less than $\lambda/2$ to guarantee the full recovery of path responses with arbitrarily high angular resolution. 
\end{criterion}

Next, we analyze the impact of channel measurement region $r$ on recovering the path responses. Aforementioned, an infinite region, i.e., $r\rightarrow+\infty$, is required to fully recover the path responses with arbitrarily high angular resolution. However, the Tx-MA can only move in a confined Tx region $\mathcal{C}_{\rm t}$, i.e., $r\leq R$, which may degrade the performance of path response recovery. Hence, the following criterion holds to enhance the recovery of path responses.

\begin{criterion}\label{limited_size}
	A limited channel measurement region leads to a path response spread in the angular domain and the channel measurement region should cover the entire limited region to minimize the path response spread, i.e., $r = R$, if there is no prior information given.
\end{criterion}

Following the above two criteria, we provide five setups for MA measurement positions, i.e., ${\bf t}^{m}=\left[x_{\rm t}^{m}, y_{\rm t}^{m}\right]^{\rm T}$ and ${\bf r}^{m}=\left[x_{\rm r}^{m}, y_{\rm r}^{m}\right]^{\rm T}, m = 1,\cdots,M$, to construct measurement matrix ${\bf \Psi}$. The MA measurement positions can be either deterministic or randomized, namely deterministic-position setup and random-position setup, respectively.

\subsection{Deterministic-Position Setups}

The deterministic-position setups for MA measurements can be regarded as uniform sampling in the Tx and Rx regions. Specifically, the Tx-MA travels throughout all measurement positions in the Tx region, while for each Tx-MA measurement position, the Rx-MA travels throughout all measurement positions in the Rx region. Following this rule, three setups of deterministic positions for Tx-MA and Rx-MA measurements are defined as follows. 

\begin{figure*}[t]
	\centering
	
	\subfigure[]{
		\begin{minipage}[t]{0.3\linewidth}
			\centering
			\includegraphics[width= 4.5 cm]{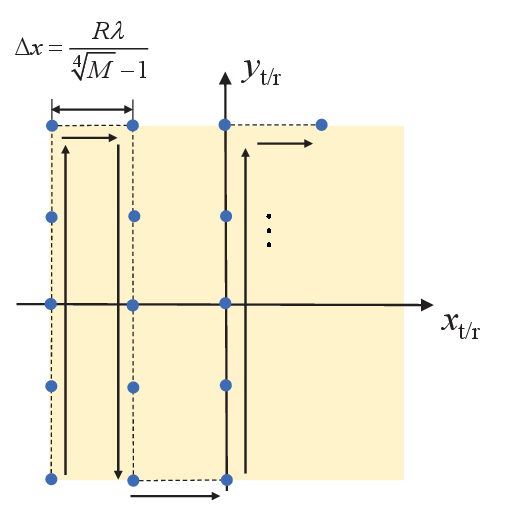}
			\label{fig:upa}
		\end{minipage}%
	}%
	\subfigure[]{
		\begin{minipage}[t]{0.3\linewidth}
			\centering
			\includegraphics[width= 4.5 cm]{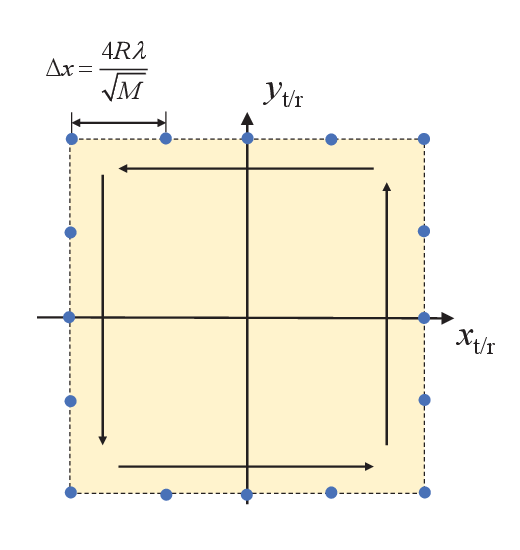}
			\label{fig:side}
		\end{minipage}%
	}
	\subfigure[]{
		\begin{minipage}[t]{0.3\linewidth}
			\centering
			\includegraphics[width= 4.5 cm]{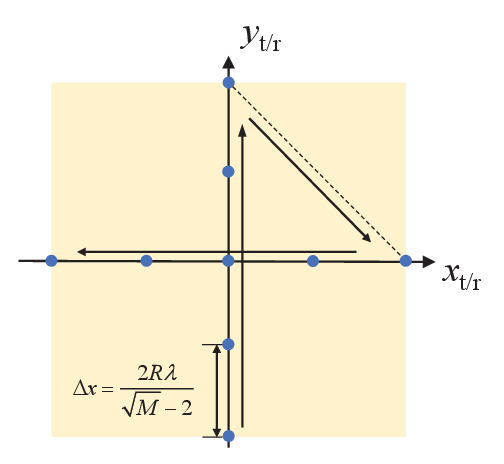}
			\label{fig:cross}
		\end{minipage}%
	}%
	
	\centering
	\caption{Illustration of the MA measurement positions in the Tx/Rx region for (a) UPA-shape setup, (b) edge of region setup, (c) cross-shape setup.}
\end{figure*}

\subsubsection{UPA-Shape}

In this setup, the positions of MAs for channel measurements are similar to uniform planar arrays (UPAs), which can be regarded as sampling in the entire Tx and Rx regions, as shown in Fig. \ref{fig:upa}. Specifically, the set of positions for MA measurements is given by
\begin{equation}
	\begin{small}
		\mathcal{P} = \left\{w_{\rm UPA}^{i} = \frac{R\lambda}{2}-i\Delta x_{\rm UPA}\left|i = 0,1,\cdots,\frac{R\lambda}{\Delta x_{\rm UPA}}\right.\right\},
	\end{small}
\end{equation}
with $\Delta x_{\rm UPA} = \frac{R\lambda}{\sqrt[4]{M}-1}$ and $M$ is the number of channel measurements. Then, for the $m$-th channel measurement, $1\leq m\leq M$, the Rx-MA and Tx-MA measurement positions are given by  
\begin{equation}
	\begin{aligned}
		x_{\rm r}^{m} = &\left(-1\right)^{\bigl\lceil \frac{m}{\sqrt{M}}\bigr\rceil}w_{\rm UPA}^{\bigl\lfloor \frac{\left(m-1\right)\bmod \sqrt{M}}{\sqrt[4]{M}}\bigr\rfloor},~x_{\rm t}^{m} = x_{\rm r}^{\tilde{m}},\\
		y_{\rm r}^{m} = &\left(-1\right)^{\bigl\lceil \frac{m}{\sqrt[4]{M}}\bigr\rceil}w_{\rm UPA}^{\left(m-1\right)\bmod \sqrt[4]{M}},~y_{\rm t}^{m} = y_{\rm r}^{\tilde{m}}.
	\end{aligned}
\end{equation}
with $\tilde{m} = \lceil \frac{m}{\sqrt{M}}\rceil$, indicating that the Rx-MA travels throughout all measurement positions for each given Tx-MA measurement position.

\subsubsection{Edge of Region}

This setup can be regarded as uniform sampling at the edges of the Tx and Rx regions, as shown in Fig. \ref{fig:side}. The set of positions for MA measurements is given by
\begin{equation}
	\begin{small}
		\mathcal{P} = \left\{w_{\rm Edge}^{i} = \frac{R\lambda}{2}-i\Delta x_{\rm Edge}\left|i = 0,1,\cdots,\frac{R\lambda}{\Delta x_{\rm Edge}}\right.\right\},
	\end{small}
\end{equation}
with $\Delta x_{\rm Edge} = \frac{4R\lambda}{\sqrt{M}}$. Then, for the $m$-th channel measurement, $1\leq m\leq M$, the Rx-MA and Tx-MA measurement positions are given by  
\begin{equation}
	\begin{aligned}
		x_{\rm r}^{m} = &\left(-1\right)^{\bigl\lceil \frac{2m}{\sqrt{M}}\bigr\rceil}w_{\rm Edge}^{\bigl\lfloor \frac{a_{m}}{\frac{\sqrt{M}}{4}+1}\bigr\rfloor \left(a_{m}-\frac{\sqrt{M}}{4}\right)},&x_{\rm t}^{m} = x_{\rm r}^{\tilde{m}},\\
		y_{\rm r}^{m} = &\left(-1\right)^{\bigl\lceil \frac{2m}{\sqrt{M}}\bigr\rceil}w_{\rm Edge}^{\left(1-\bigl\lfloor \frac{a_{m}}{\frac{\sqrt{M}}{4}+1}\bigr\rfloor\right) \left(\frac{\sqrt{M}}{4}-a_{m}\right)},&y_{\rm t}^{m} = y_{\rm r}^{\tilde{m}},
	\end{aligned}
\end{equation}
with $a_{m} = \left(m-1\right)\bmod \frac{\sqrt{M}}{2}$ and $\tilde{m} = \lceil \frac{m}{\sqrt{M}}\rceil$.

\subsubsection{Cross-shape}
This setup can be regarded as uniform sampling along the coordinate axes of the Tx and Rx regions, as shown in Fig. \ref{fig:cross}. The set of positions for MA measurements is given by
\begin{equation}
	\begin{small}
		\mathcal{P} = \left\{w_{\rm Cross}^{i} = -\frac{R\lambda}{2}+i\Delta x_{\rm Cross}\left|i = 0,1,\cdots,\frac{R\lambda}{\Delta x_{\rm Cross}}\right.\right\},
	\end{small}
\end{equation}
with $\Delta x_{\rm Cross}=\frac{2R\lambda}{\sqrt{M}-2}$. Then, for the $m$-th channel measurement, $1\leq m\leq M$, the Rx-MA and Tx-MA measurement positions are given by  
\begin{equation}
	\begin{aligned}
		x_{\rm r}^{m} =&\frac{\left(-1\right)^{\lceil \frac{2m}{\sqrt{M}}\rceil}+\left(-1\right)^{\lfloor \frac{m-1}{\sqrt{M}}\rfloor}}{2}w_{\rm Cross}^{b_{m}},&x_{\rm t}^{m} = x_{\rm r}^{\tilde{m}},\\
		y_{\rm r}^{m} =&\frac{\left(-1\right)^{\lfloor \frac{m-1}{\sqrt{M}}\rfloor}-\left(-1\right)^{\lceil \frac{2m}{\sqrt{M}}\rceil}}{2}w_{\rm Cross}^{b_{m}},&y_{\rm t}^{m} = y_{\rm r}^{\tilde{m}},
	\end{aligned}
\end{equation}
with $b_{m} = \left(\frac{\sqrt{M}}{2}-1\right)-\left(m-1\right)\bmod \frac{\sqrt{M}}{2}$ and $\tilde{m} = \lceil \frac{m}{\sqrt{M}}\rceil$.

\subsection{Random-Position Setups}

In addition to the deterministic-position setups with predefined MA measurement positions, low mutual coherence can also be achieved by introducing randomness to the MA measurement positions, which is similar to the idea of random matrices in compressed sensing theory. In the following, two setups of random positions for Tx-MA and Rx-MA measurements are defined.

\subsubsection{Random Distribution}

In this setup, the positions of MAs for channel measurements in the Tx and Rx regions follow a two-dimensional uniform distribution. Specifically, for the $m$-th channel measurement, $1\leq m \leq M$, positions of the Tx-MA and Rx-MA can be generated by
\begin{equation}
	\begin{small}
		\begin{aligned}
		x_{\rm t}^{m}&\sim\mathcal{U}\left[-\frac{R\lambda}{2},\frac{R\lambda}{2}\right],~
		y_{\rm t}^{m}\sim\mathcal{U}\left[-\frac{R\lambda}{2},\frac{R\lambda}{2}\right],\\
		x_{\rm r}^{m}&\sim\mathcal{U}\left[-\frac{R\lambda}{2},\frac{R\lambda}{2}\right],~
		y_{\rm r}^{m}\sim\mathcal{U}\left[-\frac{R\lambda}{2},\frac{R\lambda}{2}\right].
	\end{aligned}
	\end{small}
\end{equation}

\subsubsection{Random Walk}

In the random distribution scheme, an inherent challenge lies in that the Tx-MA and Rx-MA may need to move long distances between consecutive measurement positions, thereby potentially prolonging the channel estimation process. To address this problem, we propose the random walk setup, where the moving distance of the Tx-/Rx-MA between two adjacent measurements is fixed while the moving direction is random, i.e., $\Delta x_{\rm RW} = \Vert {\bf t}^{m}-{\bf t}^{m-1}\Vert_2 = \Vert {\bf r}^{m}-{\bf r}^{m-1}\Vert_2$. For initialization, let ${\bf t}^1 = {\bf r}^1 = \left[0,0\right]^{\rm T}$ for the first channel measurement. Then, for the $m$-th channel measurement, $2\leq m\leq M$, the positions of the Tx-MA and Rx-MA can be updated by
\begin{equation}
	\begin{small}
		\begin{aligned}
		x_{\rm t}^{m} =& x_{\rm t}^{m-1}+\Delta x_{\rm RW}\cos\alpha_{\rm t}^{m},~
		y_{\rm t}^{m} = y_{\rm t}^{m-1}+\Delta x_{\rm RW}\sin\alpha_{\rm t}^{m},\\
		x_{\rm r}^{m} =& x_{\rm r}^{m-1}+\Delta x_{\rm RW}\cos\alpha_{\rm r}^{m},~
		y_{\rm r}^{m} = y_{\rm r}^{m-1}+\Delta x_{\rm RW}\sin\alpha_{\rm r}^{m},\\
	\end{aligned}
	\end{small}
\end{equation}
where $\alpha_{\rm t}^{m}\sim\mathcal{U}\left[0,2\pi\right]$ and $\alpha_{\rm r}^{m}\sim\mathcal{U}\left[0,2\pi\right]$ represent the moving directions at the $\left(m-1\right)$-th measurement positions. It is worth noting that if the Tx-MA and Rx-MA reach the boundaries of the Tx and Rx regions, they should execute a `bounce-back' action, ensuring that they remain within the feasible region. For instance, if $x_{\rm t}^{m}>R\lambda/2$, then we project $x_{\rm t}^{m}$ to 
\begin{equation}
	x_{\rm t}^{m} = R\lambda-\Delta x_{\rm RW}\cos\alpha_{\rm t}^{m}-x_{\rm t}^{m-1}.
\end{equation}

\subsection{Comparison of MA Measurement Position Setups}

\begin{figure*}[t]
	\centering
	
	\subfigure[]{
		\begin{minipage}[t]{0.3\linewidth}
			\centering
			\includegraphics[width= 5.2 cm]{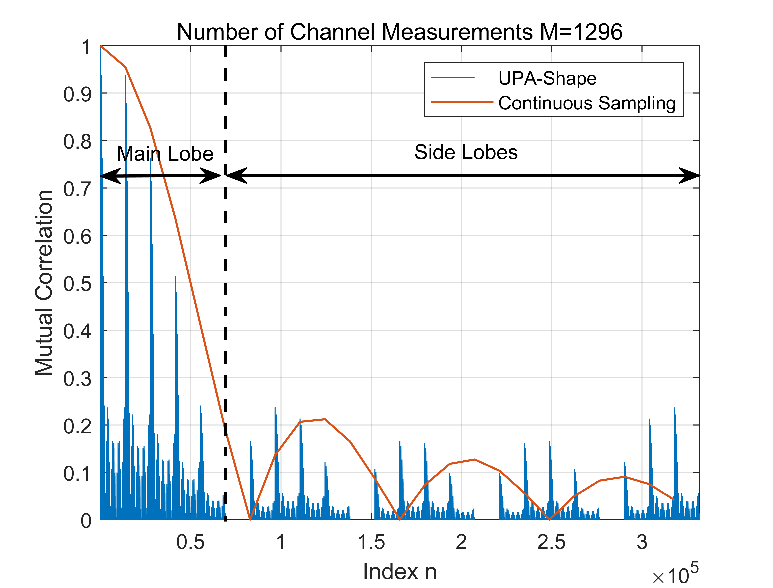}
			\label{fig:corr_upa}
		\end{minipage}%
	}%
	\subfigure[]{
		\begin{minipage}[t]{0.3\linewidth}
			\centering
			\includegraphics[width= 5.2 cm]{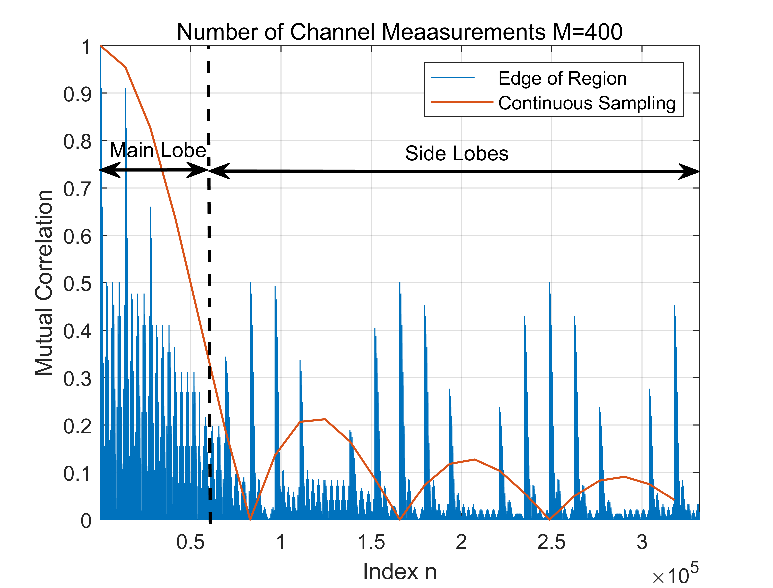}
			\label{fig:corr_side}
		\end{minipage}%
	}
	\subfigure[]{
		\begin{minipage}[t]{0.3\linewidth}
			\centering
			\includegraphics[width= 5.2 cm]{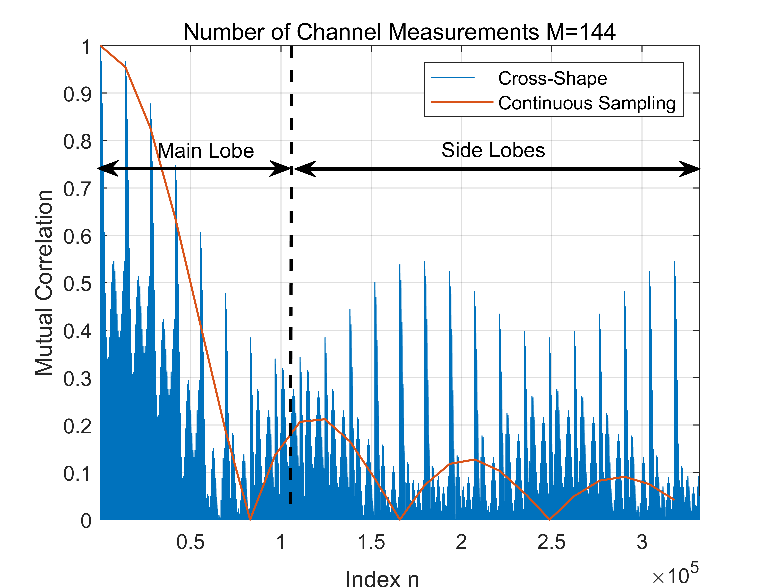}
			\label{fig:corr_cross}
		\end{minipage}%
	}%

	\subfigure[]{
	\begin{minipage}[t]{0.3\linewidth}
		\centering
		\includegraphics[width= 5.2 cm]{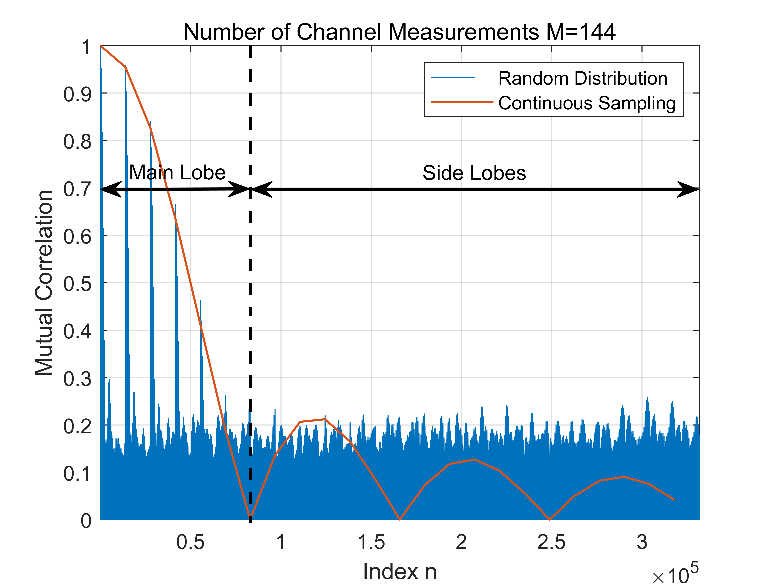}
		\label{fig:corr_random}
	\end{minipage}%
}%
\subfigure[]{
	\begin{minipage}[t]{0.3\linewidth}
		\centering
		\includegraphics[width= 5.2 cm]{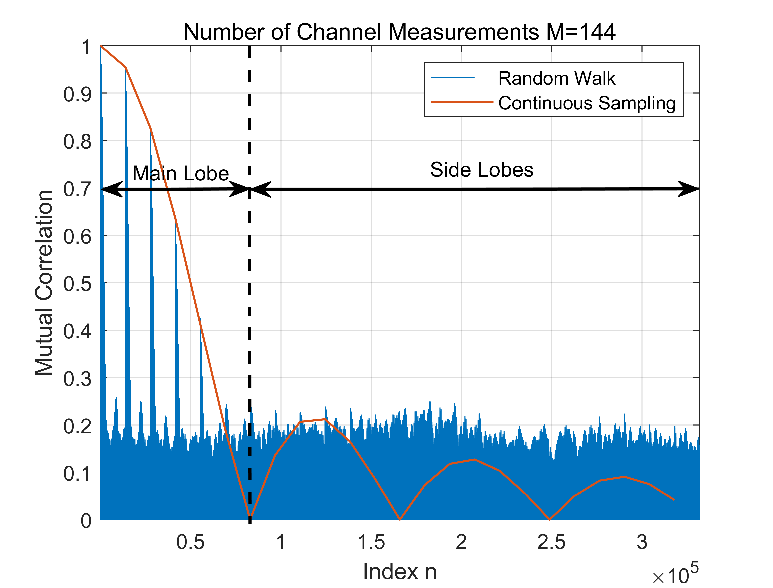}
		\label{fig:corr_randomdirection}
	\end{minipage}%
}%
	
	\centering
	\label{fig:mutual_correlation}
	\caption{Mutual coherence of proposed MA measurement position setups: (a) UPA-shape setup, (b) edge of region setup, (c) cross-shape setup, (d) random distribution setup, (e) random walk setup. }
\end{figure*}

In Fig. 4, we show the absolute value of the measurement matrix mutual coherence constructed by the proposed five MA measurement position setups with $n^{\prime}=1$ (i.e., $n_{{\rm t}x}^{\prime}=n_{{\rm t}y}^{\prime}=n_{{\rm r}x}^{\prime}=n_{{\rm r}y}^{\prime}=1$) and varying $n$ in \eqref{correlation}, i.e., absolute value of $\left[{\bf C}\right]_{:,1}=\frac{1}{M}\left[{\bf \Psi}^{\rm H}{\bf \Psi}\right]_{:,1}$. The number of quantization angles is set to $N=24$. The sizes of the Tx and Rx regions are set to $R=2$, i.e., $\mathcal{C}_{\rm t}, \mathcal{C}_{\rm r}=\left[-\lambda, \lambda\right]\times\left[-\lambda, \lambda\right]$, where $\lambda$ represents the wavelength. The spacings between adjacent MA measurement positions for the deterministic-position setups, i.e., $\Delta x_{\rm UPA}, \Delta x_{\rm Edge}$, and $\Delta x_{\rm Cross}$, are set to $0.4\lambda$ according to Criterion \ref{uniform_sampling}. Thus, the numbers of channel measurements for the UPA-shape setup, edge of region setup, and cross-shape setup are $1296, 400$, and $144$, respectively. For the random walk setup, the moving distance between two adjacent measurements is set to $\Delta x_{\rm RW}=0.5\lambda$ for both Tx-MA and Rx-MA. Moreover, the numbers of channel measurements for the random-position setups, i.e., the random distribution setup and the random walk setup, are both set to $144$, which is the same as the cross-shape setup. In addition, we draw the ideal mutual coherence under the condition of continuous sampling in the confined regions obtained in Criterion \ref{limited_size} with finite angular resolution $N$, i.e., 

\begin{equation}\label{correlation_3}
\begin{small}
	\left[\tilde{{\bf C}}\right]_{:,1} = {\rm sinc}\left(2\pi\frac{R\left(\frac{n-1}{N^3}\right)}{N}\right),
\end{small}
\end{equation}
\noindent where $n = pN^3+1, p = 1,\cdots,N-1$.

As can be observed, each mutual coherence of all these five MA measurement position setups reaches its maximum value of $1$ only when $n = n^{\prime}=1$. This indicates that Criterion \ref{uniform_sampling} is satisfied when the spacing between adjacent MA measurement positions is less than half wavelength. Besides, a main lobe and several side lobes can be observed in the mutual coherences of all the five setups due to the path response spread in the angular domain. According to Criterion \ref{limited_size}, the path response spread is caused by the limited sizes of the Tx and Rx regions. This indicates that the MA measurement positions should be distributed over the entire Tx/Rx region rather than be gathered closely.

It can be observed that compared to the deterministic-position setups, the mutual coherences achieved by the random-position setups are closer to the mutual coherence function under ideal continuous sampling. This indicates that the random-position setups can acquire more accurate channel information in the angular domain, and thereby surpass the deterministic-position setups in terms of channel estimation performance. For the deterministic-position setups, it can be observed that the MA measurement positions also have an impact on the mutual coherence. Specifically, the UPA-shape reaches the lowest mutual coherence with the most channel measurements by uniformly sampling the entire Tx and Rx regions. On the contrary, the cross-shape setup uniformly samples the coordinate axes of the Tx and Rx regions, which requires the fewest channel measurements. However, the low number of channel measurements leads to a high mutual coherence, and thereby may result in a worse performance of channel estimation. Besides, the edge of region setup performs uniform sampling at the edges of the Tx and Rx regions, and achieves a compromise between the number of channel measurements and mutual coherence.

\section{Simulation Results}\label{section_5}

In this section, the performance of the proposed compressed sensing-based channel estimation method for MA communication systems is evaluated and the proposed MA measurement position setups are further compared through comprehensive simulations. 

\subsection{Simulation Setup}
In the simulation, both Tx-MA and Rx-MA move flexibly in square areas of size $R=2$, i.e., $\mathcal{C}_{\rm t}, \mathcal{C}_{\rm r}= \left[-\lambda, \lambda\right]\times\left[-\lambda, \lambda\right]$, where $\lambda$ represents the wavelength. The geometry channel model is utilized, where each transmit path has only one corresponding receive path \cite{zhu2022modeling}. Assuming that there are $L$ paths between the Tx and Rx. In such a case, the PRM becomes a diagonal matrix characterized by $L$ non-zero entries, i.e., ${\bf \Sigma}={\rm diag}\left\{\sigma_{1}, \cdots, \sigma_{L}\right\}$. The path coefficients are independent and identically distributed (i.i.d.) CSCG random variables, i.e., $\sigma_{l}\sim\mathcal{CN}\left(0,\frac{1}{L}\right), l = 1,\cdots,L$. Besides, the MPCs are uniformly distributed over the half-space in front of the antenna panel, i.e., the physical AoDs and AoAs for the $l$-th path follow the probability density function of \cite{zhu2022modeling}
\begin{equation}
	\begin{aligned}
		f_{\rm AoD}\left(\theta_{\rm t}^{l}, \phi_{\rm t}^{l}\right)=\frac{\cos \theta_{\rm t}^{l}}{2\pi},~
		f_{\rm AoA}\left(\theta_{\rm r}^{l}, \phi_{\rm r}^{l}\right)=\frac{\cos \theta_{\rm r}^{l}}{2\pi}.
	\end{aligned}
\end{equation}

To measure the reliability of channel estimation, the normalized mean squared error (NMSE) is defined as 
\begin{equation}
	{\rm NMSE}=\mathbb{E}\left[\frac{\Vert {\bf H}-{\hat{{\bf H}}}\Vert_{\rm F}^{2}}{\Vert {\bf H}\Vert_{\rm F}^{2}}\right],
\end{equation}
where ${\bf H}\in\mathbb{C}^{D^2\times D^2}$ denotes the channel response matrix encompassing all channel responses across the entire Tx and Rx regions. Specifically, $D^2$ is a large value representing the number of points uniformly sampled in the Tx/Rx region, with their locations denoted as ${\bf t}_{\rm s}^{d^{x}_{\rm t},d^{y}_{\rm t}} ={ \left[-\frac{R\lambda}{2}+\frac{\left(d^{x}_{\rm t}-1\right)R\lambda}{D-1},-\frac{R\lambda}{2}+\frac{\left(d^{y}_{\rm t}-1\right)R\lambda}{D-1}\right]^{\rm T}, 1\leq d_{{\rm t}}^{x},d_{{\rm t}}^{y}\leq D,}$ and ${\bf r}_{\rm s}^{d_{{\rm r}}^{x},d_{\rm r}^{y}} = \left[-\frac{R\lambda}{2}+\frac{\left(d_{{\rm r}}^{x}-1\right)R\lambda}{D-1},-\frac{R\lambda}{2}+\frac{\left(d_{{\rm r}}^{y}-1\right)R\lambda}{D-1}\right]^{\rm T}$, $1\leq d_{{\rm r}}^{x},d_{{\rm r}}^{y} \leq D$. ${\bf H}$ contains the channel responses from all the Tx sampling points to all the Rx sampling points and $\hat{{\bf H}}$ is the estimation of ${\bf H}$.

In addition to NMSE, we extend our assessment to the accuracy of recovering the virtual AoD and virtual AoA, as well as the complex coefficient for each individual path. These metrics are denoted as angle error $e_{\rm Angle}$ and coefficient error $e_{\rm Coe}$, respectively. The angle error is defined as

\begin{equation}
	\begin{small}
	\begin{aligned}
		&e_{\rm Angle}=\\
		&\mathbb{E}\left[\sum_{l=1}^{L}\frac{\left|\varphi_{\rm t}^{l}-{\hat{\varphi}}_{\rm t}^{l}\right|^2+\left|\vartheta_{\rm t}^{l}-{\hat{\vartheta}}_{\rm t}^{l}\right|^2+\left|\varphi_{\rm r}^{l}-{\hat{\varphi}}_{\rm r}^{l}\right|^2+\left|\vartheta_{\rm r}^{l}-{\hat{\vartheta}}_{\rm r}^{l}\right|^2}{4L}\right],
	\end{aligned}
\end{small}
\end{equation}
where $\varphi_{\rm t}^{l}, \vartheta_{\rm t}^{l}, \varphi_{\rm r}^{l}, \vartheta_{\rm r}^{l}$ and $\hat{\varphi}_{\rm t}^{l}, \hat{\vartheta}_{\rm t}^{l}, \hat{\varphi}_{\rm r}^{l}, \hat{\vartheta}_{\rm r}^{l}$ denote the actual and estimated virtual AoDs and AoAs of the $l$-th path, respectively. On the other hand, the coefficient error measures the NMSE of the complex coefficients, i.e., 
\begin{equation}
	e_{\rm Coe }=\mathbb{E}\left[\frac{\sum_{l=1}^{L}\left|\sigma_{l}-\hat{\sigma}_{l}\right|^2}{\sum_{l=1}^{L}\left|\sigma_{l}\right|^2}\right],
\end{equation}
where $\sigma_{l}$ and $\hat{\sigma}_{l}$ represent the actual and estimated coefficients of the $l$-th path, respectively.

Unless otherwise specified, the average SNR is set to ${p_{\rm t}}/{\delta^2} = 20~\rm dB$, the number of quantization angles is set to $N=24$, and the number of paths is set to $L = 3$. The Tx-MA and Rx-MA in the random walk setup move a fixed distance of $\lambda/2$ between adjacent channel measurements, where the wavelength $\lambda$ is set to $0.01~ \rm m$. Parameter $\epsilon_0$ in \eqref{cs_channel_estimation} is set to $0.1$. To construct ${\bf H}$ and $\hat{{\bf H}}$, the number of sampling points in the Tx/Rx region is set to $D^2 = 2601$. The results in this section are obtained by $10^3$ Monte Carlo simulations. Moreover, the STRCS method proposed in \cite{10236898} is used as a benchmark for performance comparison. In particular, the UPA-shape MA position setup is used for the AoA/AoD estimation steps, and the MAs' positions are optimized via the interior-point method for the PRM estimation step.

\subsection{Numerical Results}

\begin{figure}[tbp]
	\centering
	\includegraphics[width=7.2 cm]{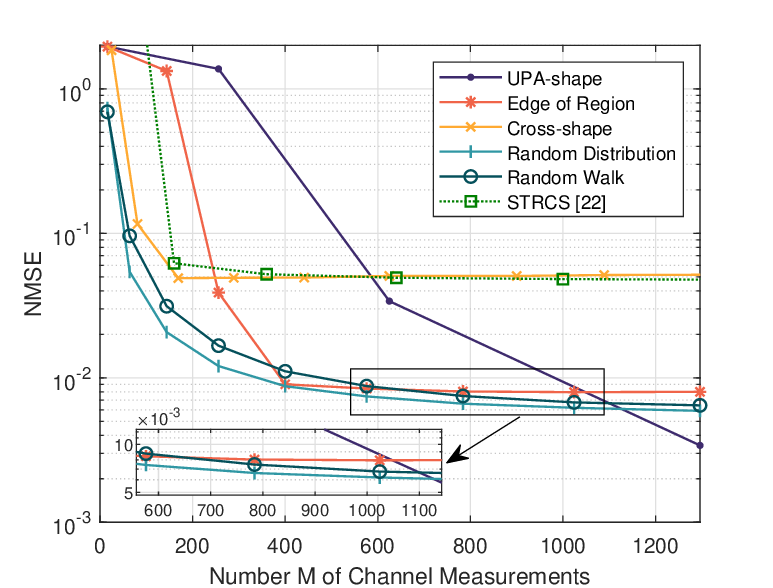}
	\caption{Comparison of the NMSE for different MA measurement position setups versus the number $M$ of channel measurements.}
	\label{fig:m_nmse}
\end{figure}

Fig. \ref{fig:m_nmse} shows the NMSE for different MA measurement position setups with varying number of channel measurements. As can be observed, random-position setups can achieve a better performance compared to deterministic-position setups, this is because the random-position setups allow the Tx-MA and Rx-MA to traverse the entire regions with a small mutual coherence of the measurement matrix. Moreover, for the random-position setups, the performance of channel estimation improves gradually as the number of channel measurements increases. This indicates that with the increasing number of channel measurements, the MPCs' information collected by channel measurement also increases. For the deterministic-position setups, since Criterion \ref{uniform_sampling} cannot be satisfied with few channel measurements, the virtual AoDs and AoAs are difficult to recover, resulting in high estimation errors. When Criterion \ref{uniform_sampling} is satisfied with a sufficiently large number of channel measurements, the NMSE of the deterministic-position setups reaches their lower bound. In other words, the NMSE will not decrease as the number of channel measurements increases to a certain level. This indicates that increasing the number of channel measurements only leads to the increase of redundancy without introducing additional MPCs' information. Moreover, it can be observed that the UPA-shape setup performs channel measurements in the entire Tx and Rx regions. Thus, it reaches the lowest lower bound on NMSE with the most channel measurements. On the contrary, the cross-shape setup performs channel measurements at the coordinate axes of the Tx and Rx regions. Thus, a small number of channel measurements is required but with a high mutual coherence, leading to a high NMSE. The edge of region setup is a compromise between the UPA-shape setup and the cross-shape setup. It can be observed that for the STRCS method, Criterion \ref{uniform_sampling} is not satisfied for a small number of channel measurements, and thus the AoD and AoA estimations are infeasible, leading to a large NMSE. When Criterion \ref{uniform_sampling} is satisfied, the STRCS method still suffers from high channel estimation error due to the sequential estimation of the AoDs, AoAs, and PRM.

\begin{figure}[tbp]
	\centering
	\includegraphics[width=7.2 cm]{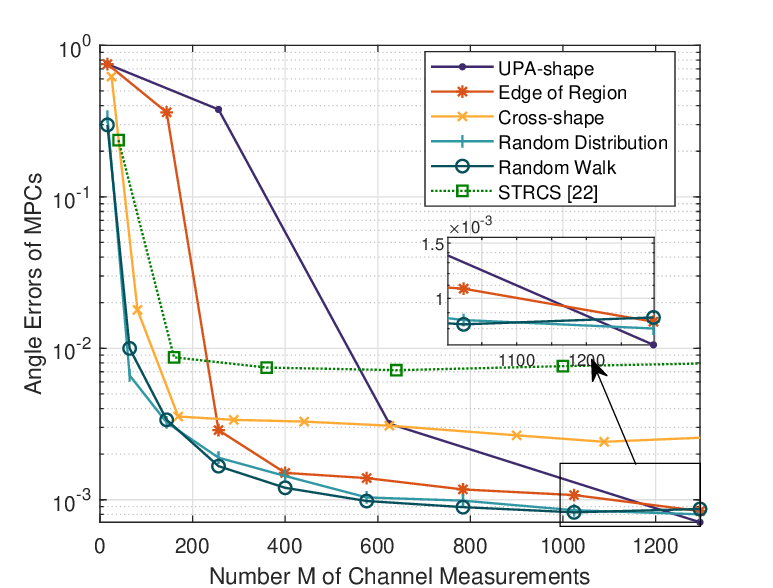}
	\caption{Comparison of the angle errors for different MA measurement position setups versus the number $M$ of channel measurements.}
	\label{fig:m_angle}
\end{figure}

In Fig. \ref{fig:m_angle}, we compare angle errors for different setups with varying number of channel measurements under the same parameter setup as Fig. \ref{fig:m_nmse}. It can be observed again that when Criterion \ref{uniform_sampling} is not satisfied, the virtual AoDs and AoAs cannot be recovered in general. Moreover, the cross-shape setup suffers the highest angle error. This is because the cross-shape setup has the highest mutual coherence as shown in Fig. \ref{fig:corr_cross}, making it difficult to accurately recover the virtual angles. Similarly, the edge of region setup also suffers a high angle error due to the high mutual coherence. The random-position setups and the UPA-shape setup reach similar performance of angle recovery due to the low mutual coherences of the three setups, while the latter one requires a large number of channel measurements.

\begin{figure}[tbp]
	\centering
	\includegraphics[width=7.2 cm]{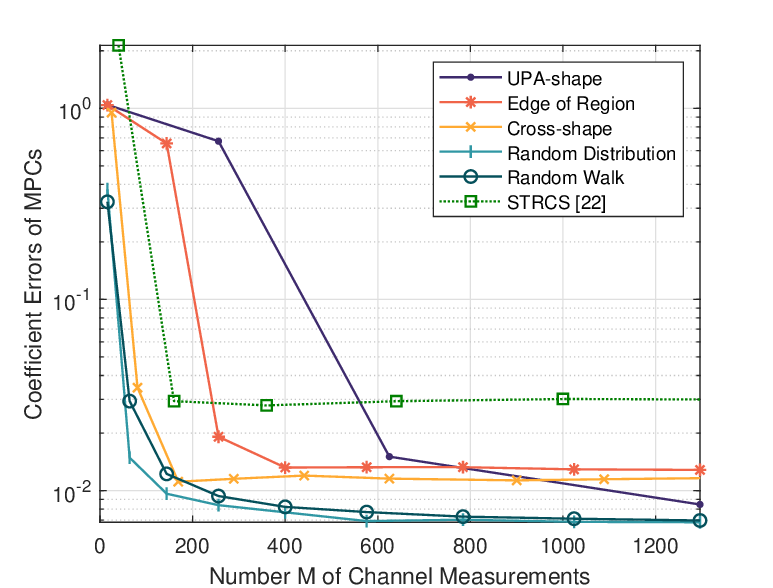}
	\caption{Comparison of the coefficient errors of MPCs for different MA measurement position setups versus the number $M$ of channel measurements.}
	\label{fig:m_gain}
\end{figure}

Furthermore, in Fig. \ref{fig:m_gain}, we compare the coefficient errors of MPCs for different MA measurement position setups with varying number of channel measurements under the same parameter setup as Fig. \ref{fig:m_nmse}. For the deterministic-position setups, when Criterion \ref{uniform_sampling} is not satisfied, the large angle error leads to a large channel coefficient error, making these MA measurement position setups less feasible for channel estimation. When the number of channel measurements is large enough, coefficient error decreases slowly with the increase of channel measurements. This is because the channel coefficients are recovered through the measured channel responses, which contain noise and quantization errors. The quantization errors exist due to the mismatch between the actual and quantized angles, which cannot be completely eliminated by increasing the number of channel measurements. For the edge of region setup, the MA measurement positions are distributed at the edge of the Tx and Rx regions, resulting in a large quantization error in the measured channel responses. The large quantization error leads to a large error in the recovered channel complex coefficients. For the cross-shape setup, the coefficient error is also high due to the incorrect estimation of the virtual angles. In addition, from the UPA-shape setup and the random-position setups, it can be observed that the coefficient error can be significantly reduced by performing channel measurements in the entire Tx and Rx regions rather than specific sub-regions (e.g., the edges of the regions or the coordinate axes of the regions).

\begin{figure}[tbp]
	\centering
	\includegraphics[width=7.2 cm]{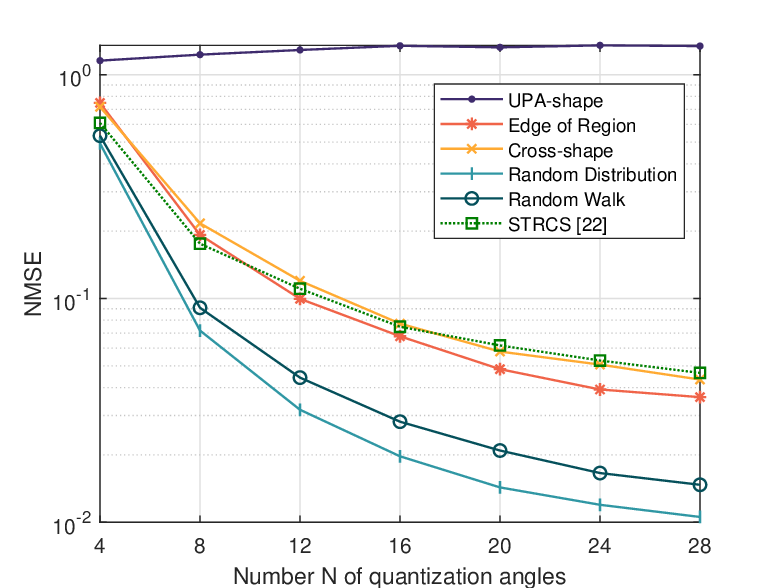}
	\caption{Comparison of the NMSE for different MA measurement position setups versus the number $N$ of quantization angles.}
	\label{fig:j_nmse}
\end{figure}

In Fig. \ref{fig:j_nmse}, we evaluate the NMSE of the MA measurement position setups with varying number $N$ of quantization angles, i.e., varying angular resolution $2/N$. The number of channel measurements is set to $M=256$ for all setups. As can be observed, for the UPA-shape setup, Criterion \ref{uniform_sampling} is not satisfied, and channel estimation is not implementable with $256$ channel measurements. For other setups, it can be observed that the NMSE significantly decreases with the increase of quantization angles, indicating that the quantization error has a significant impact on the performance of channel estimation. For the edge of region setup, the spacing between adjacent measurement positions is $\lambda/2$, indicating that the NMSE has not reached its lower bound shown in Fig. \ref{fig:m_nmse}. This demonstrates again that the deterministic-position setups require more channel measurements than the random-position setups. Besides, when $N$ is large, the NMSE of the cross-shape setup is higher than the edge of region setup, random distribution setup, and random walk setup. This is because the mutual coherence for the cross-shape setup is higher than other setups, making it more susceptible to noise and quantization errors. The performance of the cross-shape setup cannot be fully enhanced with the increase of angular resolution. According to Criterion \ref{limited_size}, it can be expected that the mutual coherence can be further reduced by enlarging the Tx and Rx regions, i.e., reducing the path response spread. Then, the performance of channel estimation can be further enhanced with a higher angle resolution. It can be observed that the STRCS method can only achieve a comparable performance to the cross-shape setup.

\begin{figure}[tbp]
	\centering
	\includegraphics[width=7.2 cm]{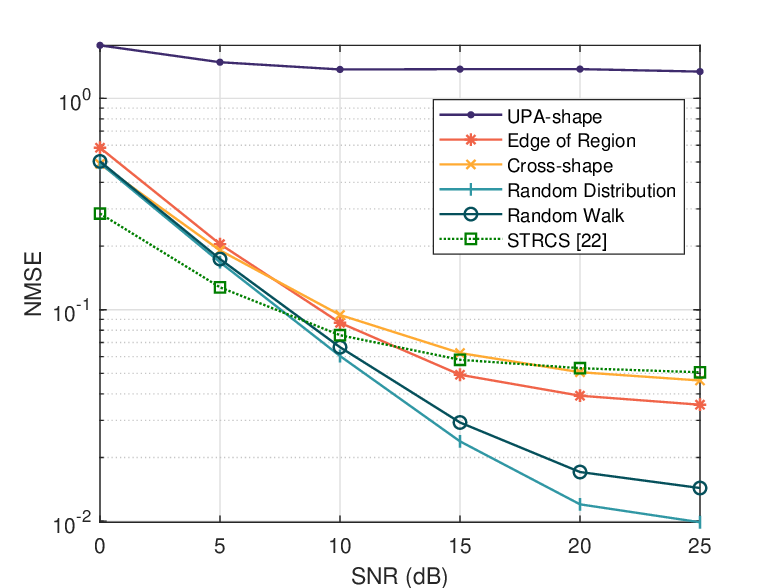}
	\caption{Comparison of the NMSE for different MA measurement position setups versus SNR.}
	\label{fig:snr_nmse}
\end{figure}

Fig. \ref{fig:snr_nmse} compares the NMSE of different setups with varying SNR. The number of channel measurements is set to $M=256$. Channel estimation for the UPA-shape setup is unsuccessful because of the large mutual coherence. For other setups, the NMSE decreases with the SNR. It can be expected that the NMSE of the edge of region setup can be reduced by shortening the spacing between adjacent measurement positions. Moreover, in Figs. \ref{fig:j_nmse} and \ref{fig:snr_nmse}, it can be observed that the random-position setups outperform the deterministic-position setups with a small number of channel measurements. Besides, the NMSE decreases slowly with the increase of SNR when the SNR is high. This indicates that the NMSE of the four setups is mainly affected by the quantization error. The STRCS method achieves a lower NMSE compared to our proposed method in the low SNR regime, i.e., ${\rm SNR} = 0~{\rm dB}$, but still infeasible for channel estimation due to the high estimation error. In the high SNR regime, the NMSE of the STRCS method is much higher than that of the proposed methods based on random MA measurement setups.

\begin{figure}[tbp]
	\centering
	\includegraphics[width=7.2 cm]{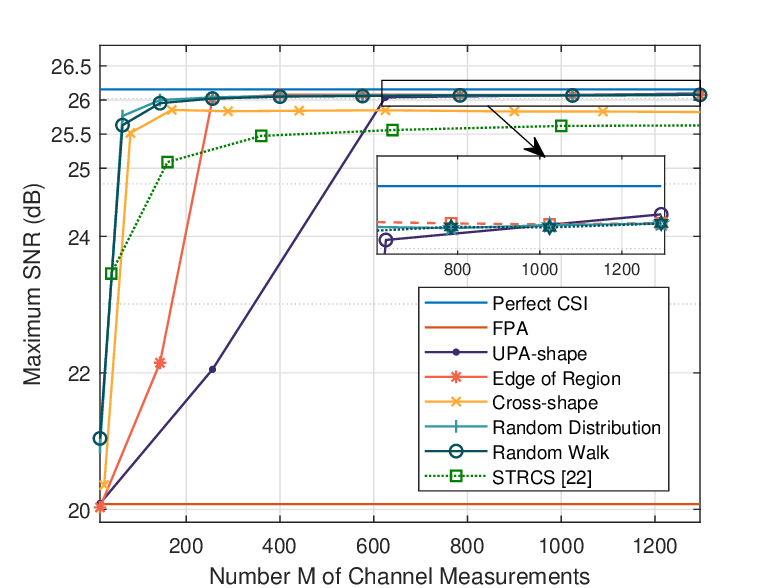}
	\caption{Comparison of the maximum SNR achieved by MA position optimization under the estimated CSI for different MA measurement position setups.}
	\label{fig:m_snr}
\end{figure}

With perfect CSI, MA systems can achieve a high SNR by moving Tx-MA and Rx-MA in the Tx and Rx regions with the best channel condition. However, the highest SNR may not always be obtained with estimated CSI due to the error of channel estimation. To evaluate this impact, in Fig. \ref{fig:m_snr}, we compare the maximum receive SNR achieved by MA position optimization under the estimated CSI obtained by different MA measurement position setups, while the maximum SNR achieved by MA position optimization under perfect CSI is set as a benchmark. In addition, we set the FPA system, whose positions of antennas are fixed at reference points, i.e., ${\bf t}^{0}$ and ${\bf r}^{0}$, as a benchmark scheme. The number of channel paths is set to $L = 5$. For FPA systems, the receive SNR is small because the antennas are fixed at reference points. On the contrary, the antennas in the MA system can move flexibly to obtain a high SNR. With perfect CSI, the receive SNR reaches the maximum value, which can be regarded as an upper bound. Consistent with the curve of NMSE in Fig. \ref{fig:m_nmse}, the random distribution setup reaches a high SNR with few channel measurements because the channel estimation error is small. Likewise, the random walk setup can reach a comparable performance to the random distribution setup. For the deterministic-position setups, more channel measurements are required to achieve a high SNR. In addition, the STRCS method can achieve a lower SNR compared to our proposed method.

\section{Conclusion}\label{section_6}

In this paper, we proposed a general channel estimation framework for MA communication systems, which can reconstruct the complete CSI between the Tx and Rx regions via a limited number of channel measurements. Specifically, based on the field-response channel model, we formulated a sparse signal recovery problem, in which the AoDs, AoAs, and complex coefficients were jointly estimated by employing the compressed sensing method. Notably, the measurement matrix for compressed sensing is determined by the Tx-MA and Rx-MA measurement positions under this framework. Moreover, we analyzed the mutual coherence of the measurement matrix from the perspective of Fourier transform, in which two criteria for MA measurement positions were obtained. Then, five MA measurement position setups for channel estimation were proposed. Finally, simulation results demonstrated that under the proposed channel estimation framework, the complete CSI between the Tx and Rx regions can be estimated with a high accuracy, and random MA measurement positions outperform deterministic measurement positions in terms of channel estimation error.

\bibliographystyle{IEEEtran} 
\bibliography{reference}

\end{document}